\documentclass[11pt]{article}
\usepackage[top=1in, bottom=1.25in, left=1.25in, right=1.25in]{geometry}
\usepackage{amsmath, amsthm, amsfonts, amssymb, amscd, bm, enumerate}

\usepackage{amsmath, amsfonts, amssymb, mathrsfs}

\usepackage{bm} 
\usepackage{cool} 
\usepackage{mathtools}
\usepackage{cases}
\usepackage{bbm}

\counterwithin{equation}{section}
\makeatletter 
\providecommand*{\diff}%
	{\@ifnextchar^{\DIfF}{\DIfF^{}}}
\def\DIfF^#1{%
	\mathop{\mathrm{\mathstrut d}}%
	\nolimits^{#1}\gobblespace}
\def\gobblespace{%
	\futurelet\diffarg\opspace}
\def\opspace{%
	\let\DiffSpace\!%
	\ifx\diffarg(%
		\let\DiffSpace\relax
	\else
		\ifx\diffarg[%
			\let\DiffSpace\relax
		\else
			\ifx\diffarg\{%
				\let\DiffSpace\relax
			\fi\fi\fi\DiffSpace}

\usepackage{xcolor}
\definecolor{oxford_blue}{RGB}{14,31,71}
\usepackage{graphicx}
\usepackage{wrapfig}
\usepackage{booktabs}
\usepackage{multirow}
\usepackage{tablefootnote}
\usepackage{multicol}
\usepackage{cite}
\usepackage[numbers]{natbib}
\usepackage[labelfont=bf]{caption} 
\usepackage{subcaption} 
\usepackage{enumerate, enumitem}
\usepackage[colorlinks]{hyperref}
\usepackage{etoolbox} 
\usepackage{url}
\usepackage{textcomp} 

\theoremstyle{remark}
\newtheorem{remark}{Remark}

\begin{document}

\title{
Estimating risks of option books using neural-SDE market models
}

\author{Samuel N. Cohen \and Christoph Reisinger \and Sheng Wang \and
Mathematical Institute, University of Oxford \\
\texttt{ \{samuel.cohen, christoph.reisinger, sheng.wang\} }\\ \texttt{@maths.ox.ac.uk}
}

\maketitle

\begin{abstract}
In this paper, we examine the capacity of an arbitrage-free neural-SDE market model 
to produce realistic scenarios for the joint dynamics of multiple European options on a single underlying. We subsequently demonstrate its use as a risk simulation engine for option portfolios. Through backtesting analysis, we show that our models are more computationally efficient and accurate for evaluating the Value-at-Risk (VaR) of option portfolios, with better coverage performance and less procyclicality than standard filtered historical simulation approaches.
\end{abstract}

{\bf MSC}: 91B28; 91B70; 62M45; 62P05

{\bf Keywords}: Market models; European options; risk measures; market simulators; no-arbitrage; neural SDE

\section{Introduction}

Managing option trade books usually requires modelling and simulating dynamics of liquid vanilla options. A desirable model, and therefore trajectories simulated from the model, should preserve statistical properties of the real data, while respecting underlying financial constraints. Previous work \cite{cohen2021mktmdl} has considered building arbitrage-free neural-SDE market models under the real-world measure, and established corresponding model estimation algorithms that take as input time series price data of finitely many liquid options. The model consists of an SDE system for states representing the underlying and a small number of factors chosen for a statistically accurate dynamic representation of the option surface and the minimisation of dynamic and static arbitrage. The coefficients of the SDE are calibrated under hard linear inequality constraints on the joint options price processes  dictated by static arbitrage relationships. In terms of practical applications, \cite{cohen2021mktmdl} only considers synthetic data simulated from a Heston stochastic local volatility model, and does not demonstrate performance on any concrete tasks. This paper validates the modelling approach in \cite{cohen2021mktmdl} using real world data, and explores the capacity of the model as a realistic option \textit{market simulator}. In particular, we shall see that these methods yield computationally efficient and accurate models for \textit{evaluating Value-at-Risk}, with less procyclicality and better performance than traditional filtered historical simulation \cite{barone2002backtesting}.

Essentially, we investigate the feasibility of using a neural-SDE market model as a \textit{generative model} for time series of option prices. Simulating synthetic data from a generative model is an important measure to ensure data privacy \cite{bellovin2019}, and can be useful for boosting deep learning performance, for instance to provide an environment to train further models with reinforcement learning \cite{buehler2019, ritter2019}. Synthetic data generators are probabilistic models which simulate observations similar to historical data, in the sense of capturing specific features of the time series. Recent developments of generative models for financial time series focus on generative adversarial networks (GANs) \cite{phl2019, ni2020conditional, koshiyama2021, takahashi2021, wiese2019deep} and variational autoencoders (VAEs) \cite{buehler2020datadriven,wiese2021multi}, and typically are not driven by fundamental financial modelling principles. We highlight the work of \cite{wiese2021multi} and \cite{Chataigner2020}, which take similar approaches to ours, using alternative representations of arbitrage conditions.

Building generative models for multiple option price time series, with the same underlying asset and different strikes and maturities, is by nature a high-dimensional problem, where complex arbitrage constraints need to be respected among different dimensions. Wiese et al. \cite{wiese2019deep,wiese2021multi} first transform option prices to an equivalent representation in the form of discrete local volatilities \cite{Wissel2008} with simple static arbitrage constraints; they then apply PCA to reduce the dimension and formulate a typical GAN problem. Compared with GAN approaches, our neural-SDE market models largely reduce the ``black-box'' nature of neural network models (see the discussion by \cite{cohen2021blackbox}) while still retaining their computational advantages. Specifically, our neural-SDE market model gives drift and diffusion estimates of each risk factor modelled, which helps us to understand the expected return and volatility of investing in each risk factor.

Here, we verify that these neural-SDE market models produce price paths similar to historical data for European equity index markets, in that they display volatility clustering (in the returns on the underlying index, and this is replicated in an options portfolio similar to VIX) and they match marginal and joint densities for the underlying and option portfolios. We then focus on risk simulation for option portfolios, which is especially convenient because the models are built under the real-world measure. In particular, we assess how well the models evaluate risks for a comprehensive set of option portfolios which represent typical risk profiles, such as (calendar) spreads, risk reversals and strangles. In fact, these portfolios can be viewed as statistical features of option price surfaces. We therefore verify that the neural-SDE market models are capable of simulating data similar to historical data, in the sense of capturing these highly structured while financially meaningful features.

Evaluating market risk for portfolios of nonlinear derivatives such as options is challenging from both an accuracy and speed perspective. We focus on Value-at-Risk (VaR), a risk measure promoted by J.P. Morgan RiskMetrics \cite{j1996riskmetrics} in the mid-1990s and quickly mandated by international regulators such as the US SEC and Basel committee for disclosing portfolio risk. 
Mathematically, the key is to establish a loss distribution for the evaluated portfolio. As our models generate samples from the full loss distribution of the portfolio, our approach could also be used to compute estimates of other risk measures, for example expected shortfall. We focus on VaR due to its common use for regulation in financial markets, and given the existence of standard backtesting approaches, which allow us to verify the performance of our method against standard criteria.

Parametric approaches to VaR estimation, such as the delta method \cite{j1996riskmetrics} and delta-gamma method \cite{Fallon1996CalculatingV, Britten1999}, give closed-form approximations of the loss distribution, which are fast, but rely heavily on distributional assumptions on the risk factors. In addition, El-Jahel et al.~\cite{El-Jahel1999} have shown that these methods give biased VaR estimates due to poor local, linear and quadratic, approximations of the loss. 

Conversely, non-parametric methods based on Monte Carlo estimation of the loss distribution are unbiased and consistent, and have flexible distributional assumptions, but can present a substantial computational burden.
First, a large number of samples of the risk factors (known as ``\textit{risk scenarios}'') are needed as VaR only considers the (upper) tail of the loss distribution. Second, for each risk scenario, one needs to re-evaluate the portfolio value, which is generally not available in closed-form, so computationally costly numerical procedures are needed. If Monte Carlo simulation is used for pricing the portfolio, one ends up with a two-level (or nested) simulation scheme. Variance reduction techniques can be used in both outer and inner levels to accelerate simulations; specifically, for the outer level, where quantile estimation is made, there is a vast literature discussing the use of control variates \cite{hsu1990, Hesterberg1998}, antithetic variates \cite{Avramidis1998}, and, more popularly, importance sampling or stratified sampling \cite{Glasserman2000, Glasserman2002, SUN2010}. Further non-parametric VaR estimation methods include historical simulation \cite{butler1997estimating} or volatility-filtered historical simulation \cite{barone1997var, barone2001non, barone2002backtesting, fhs2015}, which sample risk scenarios based on empirical distributions of risk factors. Nevertheless, regardless of how risk scenarios are simulated, the valuation of options portfolios often requires re-evaluation of option prices based on risk factors (for multiple simulation trajectories and time points), which is computationally expensive.

In our market models, option prices are linear in the risk factors, resulting in negligible computational cost for re-evaluating an option portfolio for each simulated risk scenario. Despite the simplicity of the linear factor representation, it is sufficiently rich to achieve satisfactory statistical accuracy. As shall be detailed later, we simulate risk factors from a flexible, offline trained neural-SDE with historical innovations. Through out-of-sample backtesting, we show that our model yields statistically and economically more appealing VaR estimates for various option portfolios than traditional filtered historical simulation approaches (see, for example, Barone-Adesi, Giannopoulos and Vosper \cite{barone2002backtesting}).

This paper is organised as follows. First, we review the construction and estimation methodology of the arbitrage-free neural-SDE market model in Section \ref{sec:review_mdl}. Then we explain the dataset used, elaborate how to apply the model to real world data in Section \ref{sec:calib_to_data}, where the in-sample and out-of-sample performance of the model is assessed extensively. Given the trained model, we estimate VaR over various risk horizons and confidence levels for a varied set of option portfolios, and backtest the out-of-sample VaR performance in Section \ref{sec:risk_mgt}. Finally, Section \ref{sec:conclusion} concludes our observations and addresses a few further extensions.

\section{Arbitrage-free neural-SDE market models}
\label{sec:review_mdl}

Here we briefly summarise our neural-SDE approach to constructing market models. Interested readers may refer to \cite{cohen2021mktmdl} for a discussion of model choices, theoretical strengths and limitations, and other implementational details.

\subsection{Factor-based market models}

We consider a financial market with a stock\footnote{Although we refer to the asset as ``stock'' throughout the paper, our methods are equally applicable to other assets such as equity indices, currencies or commodities.} and a collection of European call options on the stock. We denote by $C_t(T,K)$ the market price at time $t$ of an European call option with expiry $T$ and strike $K$. Over time, the range of liquid options in $(T,K)$-coordinates tends to change \emph{stochastically}. In contrast, there is usually a stable range of time-to-expiries and moneynesses for which the options are actively quoted and therefore price data are most readily available. Therefore, it is empirically advantageous to build models in terms of relative coordinates $(\tau, m)$, where $\tau=T-t$ is time-to-expiry and $m$ denotes moneyness. Specifically, we model the \emph{normalised call price surface} $(\tau, m) \mapsto \tilde{c}_t (\tau, m)$ with the definition
\begin{equation}
    \tilde{c}_t (\tau, m) = \frac{C_t(t+\tau, e^m F_t(t+\tau))}{D_t(t+\tau)F_t(t+\tau)},
\label{eq:call_price_transformation}
\end{equation}
where $D_t(T)$ denotes the market discount factor for time $T$ (assuming a deterministic interest rate), and $F_t(T)$ the model-independent, arbitrage-free futures price for delivery of the stock at $T$.

We fix a constant compact set $\mathcal{R}_\text{liq} \subset \{(\tau, m) \in \mathbb{R}^2 \}$, which represents the range of time-to-expiries and moneynesses where options are liquid. For each $\tau$ and $m$, we assume a time-independent linear representation of $\tilde{c}_t(\tau, m)$ by \emph{latent factors} $\xi \in \mathbb{R}^d$, given by
\begin{equation}
    \tilde{c}_t(\tau,m) = G_0(\tau, m) + \sum_{i=1}^d G_i(\tau, m) \xi_{it},
\label{eq:factr_rep_linear}
\end{equation}
where $G_i \in C^{1,2}(\mathcal{R}_\text{liq})$ for $i=0,\dots,d$, and we refer to it as a \emph{price basis function}. Thereafter, we model the joint dynamics of the stock price $S$ and the factors $\xi$ by a $(d+1)$-dimensional diffusion process solving the following time-homogeneous SDE:
\begin{subequations}
\begin{align}
    \frac{\diff S_t}{S_t} & = \alpha(\xi_t) \diff t + \gamma(\xi_t) \diff W_{0,t}, & S_0 = s_0 \in \mathbb{R}; \label{eq:market_model_S} \\
    \diff \xi_t & = \mu (\xi_t) \diff t + \sigma (\xi_t) \diff W_t,  & \xi_0 = \zeta_0 \in \mathbb{R}^d, \label{eq:market_model_xi}
\end{align}%
\label{eq:market_model}%
\end{subequations}%
where $W_{0} \in \mathbb{R}$ and $W =[W_{1} ~\cdots ~W_{d}]^\top \in \mathbb{R}^d$ are standard independent Brownian motions under the real-world measure $\mathbb{P}$. We use $\alpha_t$, $\gamma_t$, $\mu_t$ and $\sigma_t$ to denote $\alpha(\xi_t)$, $\gamma(\xi_t)$, $\mu(\xi_t)$ and $\sigma(\xi_t)$, respectively. We denote by $L^p_\text{loc}(\mathbb{R}^d)$ the space of all $\mathbb{R}^d$-valued, progressively measurable, and locally $p$-integrable (in $t$, $\mathbb{P}$-a.s.) processes, and assume that $\alpha \in L^1_\text{loc}(\mathbb{R})$, $\mu \in L^1_\text{loc}(\mathbb{R}^{d})$, $\gamma \in L^2_\text{loc}(\mathbb{R})$ and $\sigma \in L^2_\text{loc}(\mathbb{R}^{d})$.

\begin{remark}
We assume, in the model of $\xi$ \eqref{eq:market_model_xi}, that the drift and diffusion coefficients are independent of $S$, in the sense that volatility surfaces (represented by $\xi$) are unaffected by the underlying price level. Nevertheless, it is possible, and simple, to add $S$ as an additional argument and train the model with neural nets (as is done in \cite{cohen2021mktmdl}).
\end{remark}

\begin{remark}
\label{rmk:mdl_xi_S}
For simplicity, we assume independent Brownian motions that drive the randomness in $S$ and $\xi$; one can easily add a correlation structure and estimate the resulting larger covariance matrix, albeit with a greater risk of over-fitting. In Appendix \ref{apd:model_xiS}, we discuss the performance when estimating models for $S$ and $\xi$ jointly with a full $\mathbb{R}^{(d+1)\times(d+1)}$ covariance matrix. With the current model setup \eqref{eq:market_model}, in order to take into account non-negligible correlations between $\xi$ and log-returns of $S$ for realistic forward simulations, we randomly sample innovations from the historical residuals of the fitted models \eqref{eq:market_model}, rather than from standard Gaussians; see the discussion in Section \ref{sec:factor_simulation}.
\end{remark}

\subsection{Arbitrage considerations}

According to the First Fundamental Theorem of Asset Pricing (FFTAP) \cite{harrison1979}, a model admits no arbitrage if there exists a measure $\mathbb{Q} \sim \mathbb{P}$ such that the discounted price processes for all tradable assets are martingales. (Formally, this can be weakened to local or $\sigma$-martingales, depending also on the precise form of no-arbitrage assumed.) We derive an HJM-type drift restriction (see also \cite{HJM1992}) for this no-arbitrage condition to hold, by analyzing the existence of a \emph{market price of risk} process $\psi_t = [\psi_{1,t} ~\cdots ~\psi_{d,t}]^\top$. Considering a finite set of option specifications $\mathcal{L}_\text{liq} = \{ (\tau_j, m_j) \}_{j=1,\dots,N} \subset \mathcal{R}_\text{liq}$, we look for a solution $\psi_t$ to
\begin{equation}
    \mathbf{G}^\top \sigma_t \psi_t=\mathbf{G}^\top \mu_t - z_t, \quad \text{ where } z_t = \left( -\pderiv{}{\tau} - \frac{1}{2} \gamma_t^2 \pderiv{}{m} + \frac{1}{2} \gamma_t^2 \pderiv[2]{}{m} \right) \mathbf{c}_t,
    \label{eq:mpor_linear_system}
\end{equation}
with $\mathbf{c}_t = [\tilde{c}_t(\tau_1, m_1) ~\cdots ~\tilde{c}_t(\tau_N, m_N)]^\top$. Here, we define $\mathbf{G}$ as the $d\times N$ matrix with $i$-th row $\mathbf{G}_i$, where $\mathbf{G}_i = (G_{ij})_{j=1}^N \in \mathbb{R}^N$ for $i=0,\dots, d$, and $G_{ij} = G_i(\tau_j, m_j)$. We call $\mathbf{G}^\top$ a \emph{price basis matrix} and each of its columns a \emph{price basis vector}. The relation \eqref{eq:mpor_linear_system} leads to a mechanism of discouraging dynamic arbitrage in the factor decoding algorithm (Algorithm 1 of \cite{cohen2021mktmdl}), and a metric for measuring the proportion of dynamic arbitrage in factor-reconstructed prices (Section 4.3 of \cite{cohen2021mktmdl}).

Nevertheless, the HJM-type drift restrictions cannot be implemented exactly in practice, because our models are built on finitely many options on a fixed liquid lattice in the $(\tau, m)$-coordinates, which do not correspond to fixed contracts (i.e.\ fixed $T$ and $K$) over time, and so solutions to \ref{eq:mpor_linear_system} will not exist for a fixed price basis $\mathbb{G}$. Therefore, further consideration of no-arbitrage conditions is needed. In particular, we can consider static, model-free arbitrage constraints on the underlying factors. Using the construction in our previous work \cite{Cohen2020}, we write static arbitrage constraints in the form $\mathbf{A} \mathbf{c}_t \geq \widehat{\mathbf{b}}$, where $\mathbf{A} = (A_{ij}) \in \mathbb{R}^{R \times N}$ and $\widehat{\mathbf{b}} = (\hat{b}_j) \in \mathbb{R}^{R}$ are a known constant matrix and vector, and $R$ is the number of static arbitrage constraints. Given the factor representation \eqref{eq:factr_rep_linear}, we have $\mathbf{c}_t = \mathbf{G}_0 + \mathbf{G}^\top \xi_t$. Consequently, the market model allows no \emph{static} arbitrage among options on the liquid lattice $\mathcal{L}_\text{liq}$ if $\xi_t$ satisfies, for all $t$,
\begin{equation}
\mathbf{A} \mathbf{G}^\top \xi_t \geq \mathbf{b} := \widehat{\mathbf{b}} - \mathbf{A G}_0^\top.
\label{eq:factor_static_arbitrage}
\end{equation}

\subsection{Model inference}

As the first step of model inference, we decode the factor representation \eqref{eq:factr_rep_linear} by calibrating the price basis functions $\{G_i\}_{i=0}^d$. This leads to Algorithm 1 of \cite{cohen2021mktmdl}, which extracts a smaller number of market factors from prices of finitely many options with the joint objectives of eliminating static and dynamic arbitrage in reconstructed prices and guaranteeing statistical accuracy.

To estimate the drift and diffusion of the decoded factor process $\xi$, as specified in \eqref{eq:market_model_xi}, we represent the drift and diffusion functions by neural networks, referred to as \emph{neural-SDE}. Leveraging deep learning algorithms, we train the neural networks by maximising the likelihood of observing the factor paths, subject to the derived arbitrage constraints. As seen in \eqref{eq:factor_static_arbitrage}, static arbitrage constraints are characterised by the convex polytope state space $\{y \in \mathbb{R}^d:\mathbf{A} \mathbf{G}^\top y \geq \mathbf{b}\}$ for the factors; we identify sufficient conditions on the drift and diffusion to restrict the factors to the polytope, using the classic results of Friedman and Pinsky \cite{friedman1973}. Consequently, the neural network that is used to parameterise the drift and diffusion functions needs to be constrained. We achieve this by developing appropriate transformations for the output of the neural network, as detailed in Section 3 of \cite{cohen2021mktmdl}.

\section{Calibration to market data}
\label{sec:calib_to_data}

We have seen decent performance of our modelling and inference approach when applied to synthetic data generated from a stochastic local volatility model (Jex, Henderson and Wang \cite{Jex1999}). The focus here is to investigate its ability to capture the dynamics of real-world data, and whether practically useful conclusions can be drawn from these methods.

\subsection{Option price data description and pre-processing}
\label{sec:option_data}

We study the two most traded vanilla European options listed on Eurex, namely EURO STOXX 50{\small\textsuperscript\textregistered} index options and DAX{\small\textsuperscript\textregistered} options. OptionMetrics' IvyDB Europe \cite{optionmetrics} offers historical daily settlement prices\footnote{These prices are calculated from implied volatilities that are interpolated using a methodology based on a kernel smoothing algorithm, according to the OptionMetrics' IvyDB Europe reference manual \cite{optionmetrics}.} for calls with expiries of 30, 60, 91, 122, 152, 182, 273, 365, 547, and 730 calendar days, at deltas of 0.2, 0.25, 0.3, 0.35, 0.4, 0.45, 0.5, 0.55, 0.6, 0.65, 0.7, 0.75, and 0.8. We collect daily call option prices from 2nd January 2002 to 30th December 2019, as well as corresponding futures curves and euro zero curves.

\begin{figure}[!ht]
    \begin{subfigure}[b]{.31\textwidth}
    \centering
        \includegraphics[scale=.66]{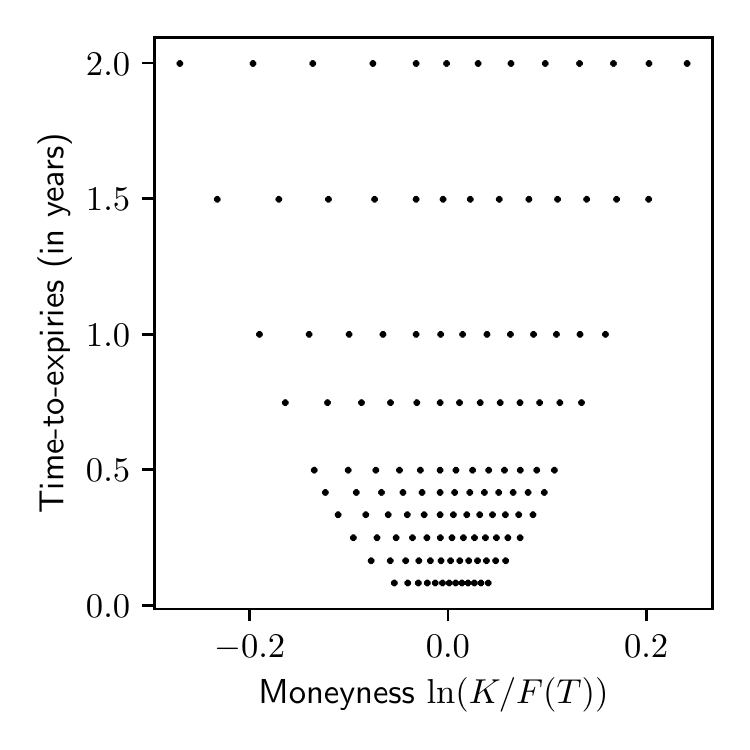}
        \caption{Liquid lattice $\mathcal{L}_\text{liq}$.}
        \label{fig:eurex_lattice}
    \end{subfigure}
    \hfill
    \begin{subfigure}[b]{.68\textwidth}
    \centering
        \includegraphics[scale=.66]{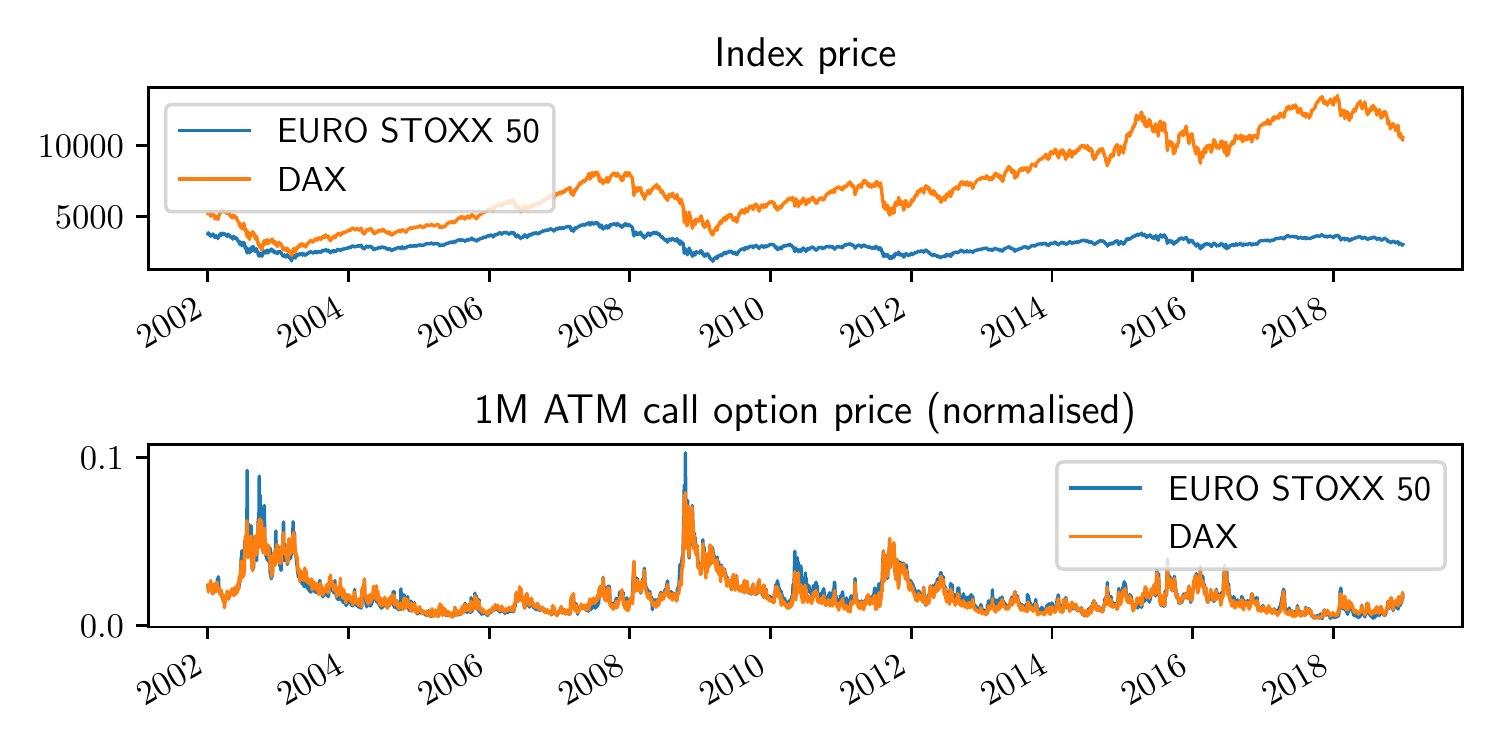}
        \caption{Historical daily prices (underlying indices and 1M ATM call options).}
        \label{fig:eurex_ts}
    \end{subfigure}
    \caption{Scattergram of liquid option lattice and historical prices.}
\end{figure}

We first normalise call prices by discount factors and futures prices following \eqref{eq:call_price_transformation}. However, the normalised call price surfaces have fixed delta parameters, rather than fixed moneynesses, over time. Therefore, we choose the historical \emph{median} moneynesses (for each delta-quoted option) to define our liquid option lattice $\mathcal{L}_\text{liq}$, and interpolate prices to the fixed set of moneyness parameters. We give details on the interpolation in Appendix \ref{apd:data_cleansing}. In addition, whenever there is static arbitrage in the reported historical option prices, we perturb the prices using the arbitrage repair algorithm of \cite{Cohen2020}. In Appendix \ref{apd:data_cleansing}, we give statistics on the presence of arbitrage in raw and interpolated option price data, and assess the impact of repairing arbitrage. We show the liquid lattice $\mathcal{L}_\text{liq}$ in Figure \ref{fig:eurex_lattice}.

In Figure \ref{fig:eurex_ts}, we show historical daily prices for the EURO STOXX 50 index, the DAX index, and the 1M at-the-money (ATM) normalised call options written on the two indices. Since the normalised option price has a deterministic bijective relation with the implied volatility, we see that the two options have very similar volatility dynamics in history. It is consequently reasonable to assume a common background model for the volatility processes of the two options. Therefore, we concatenate the normalised call price data of the EURO STOXX 50 index options and DAX options when decoding factors and learning factor dynamics. This enriches the dataset available for our statistical estimation problems by assuming that our market models are universal for options written on these two different underlyings. Specifically, there is only one functional form for the drift $\mu(\cdot)$ and the diffusion $\sigma(\cdot)$ for the factor dynamics \eqref{eq:market_model}, despite different option underlyings. Nevertheless, the underlying indices' models will not be restricted to be the same; as evidenced in Figure \ref{fig:eurex_ts}, DAX index appears to have a more positive drift\footnote{Two factors contribute to the drift difference in the two indices: in addition to the fact that the two equity indices have different constituent stocks, EURO STOXX 50 is a price return index while DAX is a total return index.}. Hence, we build a universal model for the factor dynamics but separate models for each index.

Using $k \in \{U, D\}$ to distinguish data of EURO STOXX 50 index options and DAX options, we denote the observation times as $0 = t_1 < \cdots < t_{L^k} = T^k$ and let $\delta t = t_i - t_{i-1} = T^k / (L^k-1)$ for all $2 \leq i \leq L^k$. We split the $L^k$ historical data samples into \textit{training}/\textit{estimation} samples $\{t_1, \dots, t_{E^k} \}$ and \textit{testing} samples $\{t_{E^k+1}, \dots, t_{L^k}\}$. We use only the training samples for estimating a neural-SDE market model, while the testing samples will be used for testing performance of out-of-sample risk simulation, as detailed in Section \ref{sec:bar_backtesting}. In Table \ref{tab:data}, we list the details of the data used. 

\begin{table}[!ht]
    \centering
    \footnotesize
    \begin{tabular}{ccccc}
    \toprule
        \textbf{Parameters} & \#training samples $E^k$ & \#testing samples $L^k - E^k$ & \#options $N$ & $\delta t$\\ \cmidrule(lr){1-1} \cmidrule(lr){2-5}
        \textbf{Values} & 4296 (2002-01-01 $-$ 2018-12-31) & 253 (2019-01-01 $-$ 2019-12-31) & 130 & 1 day\tablefootnote{We use the Act/365 fixed day count convention for converting $\delta t = 1/365$ to year fraction. } \\
    \bottomrule
    \end{tabular}
    \caption{Details of the option data used.}
    \label{tab:data}
\end{table}

\subsection{Factor decoding}

Suppose we observe price data $\mathbf{C}^k = (C^k_{lj}) \in \mathbb{R}^{L^k \times N}$ with $C^k_{lj} = \tilde{c}^k_{t_{l}}(\tau_j, m_j)$ for $k \in \{E, D\}$. Given the concatenated data $\mathbf{C} = [\mathbf{C}^E; \mathbf{C}^D] \in \mathbb{R}^{L \times N}$ with $L = L^E + L^D$, we aim to construct factor data $\bm\Xi = (\Xi_{li}) \in \mathbb{R}^{L \times d}$ with $\Xi_{li} = \xi_{i, t_{l}}$. In particular, we represent $\mathbf{C}$, with residuals $\bm\Upsilon \in \mathbb{R}^{L \times N}$, by
\begin{equation}
	\mathbf{C} = \mathbf{1}_L \otimes \mathbf{G}_0 + \bm\Xi \mathbf{G} + \bm\Upsilon,
\label{eq:represent_C}
\end{equation}
where $\mathbf{1}_{L}$ is an $L$-vector of ones, and $\otimes$ denotes the outer product of two vectors. 

\subsubsection*{Vega-weighted prices}

We call $\mathbf{C} - \bm\Upsilon$ the \textit{reconstructed prices}, and the Frobenius norm $||\bm\Upsilon||_F$ the \textit{reconstruction error}. As in-the-money (ITM) and long-dated options have higher prices, the objective of minimising $||\bm\Upsilon||_F$ is naturally more effective in reducing relative calibration errors for ITM and long-dated option prices. To avoid overfitting ITM and long-dated option prices, practitioners often weight each option by the reciprocal of its Black--Scholes vega. In fact, using the approximation $\Delta C \approx \text{vega} \times \Delta \sigma^\text{imp}$, weighting prices by $1/\text{vega}$ approximately corresponds to measuring reconstruction errors in implied volatilities. To follow this approach, we introduce a constant weight matrix $\bm\Lambda = \textrm{diag} \left( \lambda_1, \dots, \lambda_N \right)$, and construct factors from $\mathbf{C} \bm\Lambda$ instead, as the following two optimisation problems are equivalent:
\begin{equation*}
    \underset{\bm\Xi, \mathbf{G}}{\min} \Big\| \left( \mathbf{C} - \mathbf{1}_L \otimes \mathbf{G}_0 - \bm\Xi \mathbf{G}\right) \bm\Lambda \Big\|
    \Longleftrightarrow
    \underset{\widehat{\bm\Xi}, \widehat{\mathbf{G}}}{\min} \Big\| \mathbf{C}\bm\Lambda - \mathbf{1}_L \otimes \widehat{\mathbf{G}}_0 - \widehat{\bm\Xi}\widehat{\mathbf{G}} \Big\|.
\end{equation*}

Now we associate the weight $\lambda$ with the option vega. At time $t$, the Black--Scholes vega for the $(\tau, m)$-option is 
\begin{equation}
	\mathcal{V}_t(\tau, m) = \frac{\diff \tilde{c}_t}{\diff \sigma^\text{imp}_t} (\sigma^\text{imp}_t; \tau,m)  = \sqrt{\tau} \phi \left(-\frac{m}{\sigma^\text{imp}_t \sqrt{\tau}} + \frac{1}{2}\sigma^\text{imp}_t \sqrt{\tau} \right) > 0,
\label{eq:vega}
\end{equation}
where $\phi(\cdot)$ is the density of the standard normal distribution. This leads us to set
\begin{equation*}
    \lambda_j = \frac{1}{\bar{\mathcal{V}}_j}, \text{
    for each $j$, where } \bar{\mathcal{V}}_j = \frac{1}{L}\sum_{l=1}^L \mathcal{V}_{t_l}(\tau_j, m_j).
\end{equation*}

\begin{remark}
Weighting price data by a constant matrix does not change the factor decoding algorithm (Algorithm 1 of \cite{cohen2021mktmdl}), except that the coefficient and constant terms of the static arbitrage constraints \eqref{eq:factor_static_arbitrage} need to be revised. Specifically, we aim for a new factor representation $\bm\Lambda \mathbf{c}_t = \mathbf{G}_0 + \mathbf{G}^\top \xi_t$. Then the static arbitrage constraints $\mathbf{A} \mathbf{c}_t \geq \widehat{\mathbf{b}}$ yield
\begin{equation*}
    \left( \mathbf{A} \bm\Lambda^{-1} \right) \mathbf{G}^\top \xi_t \geq \mathbf{b}^\text{new} := \widehat{\mathbf{b}} -  \left( \mathbf{A} \bm\Lambda^{-1} \right) \mathbf{G}_0.
\end{equation*}
Nevertheless, for notational simplicity, we write $\mathbf{C} \leftarrow \mathbf{C} \bm\Lambda$, $\mathbf{A} \leftarrow \mathbf{A} \bm\Lambda^{-1}$ and $\mathbf{b} \leftarrow \mathbf{b}^\text{new}$ for the rest of the article.
\end{remark}

\subsubsection*{Primary and secondary factors}

While decoding more factors can improve the representation of call prices, it also leads to higher-dimensional models, which may over-parametrise the dynamics of noisy factors and lead to over-fitting. To overcome these issues, we assume that some \textit{primary} factors are modelled by the SDE \eqref{eq:market_model} and other \textit{secondary} factors are modelled as mean-reverting Ornstein--Uhlenbeck (OU) processes.

We now investigate how many primary and secondary factors are needed for a reasonable representation of the call price data. In Section 4.3 of \cite{cohen2021mktmdl}, we define three metrics to examine the quality of factor reconstruction, which are MAPE (mean absolute percentage error), PDA (proportion of dynamic arbitrage) and PSAS (proportion of statically arbitrageable samples)\protect\footnotemark. As discussed in the previous section, we use vega-weighted versions of these quantities in what follows. Applying the factor decoding algorithm (Algorithm 1 of \cite{cohen2021mktmdl}), we show the three metrics for a few combinations of using up to two factors in Table \ref{tab:factor_metrics}.

\footnotetext{
We recall that the MAPE and PSAS metrics are defined by
\begin{equation*}
    \text{MAPE} = \frac{1}{(L+1)N} \sum_{l=0}^L \sum_{j=1}^N \frac{\left| \tilde{c}_{t_l}(\tau_j, m_j) / \bar{\mathcal{V}}_j  - G_{0j} - \sum_{i=1}^d G_{ij} \xi_{it_l} \right|}{\tilde{c}_{t_l}(\tau_j, m_j) / \bar{\mathcal{V}}_j},
\end{equation*}
\begin{equation*}
    \text{PSAS} = 1 - \frac{ \sum_{l=0}^L \bm{1}_{\{\mathbf{A} \bm\Lambda^{-1} \mathbf{G}^\top \xi_{t_l} \geq \mathbf{b} \}}}{L+1}.
\end{equation*}
The formula for PDA is somewhat involved, we refer readers to Section 4.3 of \cite{cohen2021mktmdl}.
}

\begin{table}[!ht]
    \centering
    \footnotesize
    \begin{tabular}{lccc}
        \toprule
        Factors & MAPE & PDA & PSAS  \\
        \cmidrule(lr){1-1} \cmidrule(lr){2-4}
        Dynamic arb. & $10.75\%$ & $2.20\%$ & $33.74\%$  \\
        Dynamic arb. + Statistical acc. & $4.61\%$ & $1.92\%$ & $8.26\%$ \\
        Dynamic arb. + Static arb. & $4.94\%$ & $1.99\%$ & $1.29\%$ \\
        Statistical acc. & $5.40\%$ & $13.07\%$ & $7.13\%$ \\
        Statistical acc. + Dynamic arb. & $5.40\%$ & $1.80\%$ & $8.13\%$ \\
        Statistical acc. + Static arb. & $4.82\%$ & $7.83\%$ & $0.30\%$ \\
        \bottomrule
    \end{tabular}
    \caption{MAPE, PDA and PSAS metrics when including different combinations of factors.}
    \label{tab:factor_metrics}
\end{table}

Using only two factors, i.e.\ one statistical accuracy factor and one static arbitrage factor (last row in the table), we can represent the whole collection of call prices with reasonable accuracy. Note that the use of a static arbitrage factor is significant in reducing the number of violations of the static arbitrage constraints, as evidenced by the corresponding reduction in PSAS. Since arbitrageable data samples are invalid inputs for the model inference and need to be thrown away, it is crucial to ensure a low PSAS. The high value of PDA is of less concern for real data and will be reduced further by including more secondary factors. Hence, we will restrict our attention to this simple three-factor model (i.e.\ two factors $\xi$, in addition to the stock index price), as it also allows us to demonstrate qualitative features of the model easily. See, however, Appendix \ref{apd:3factor} for the analysis of a market model with three primary factors.

\begin{figure}[!ht]
    \centering
    \includegraphics[scale=.66]{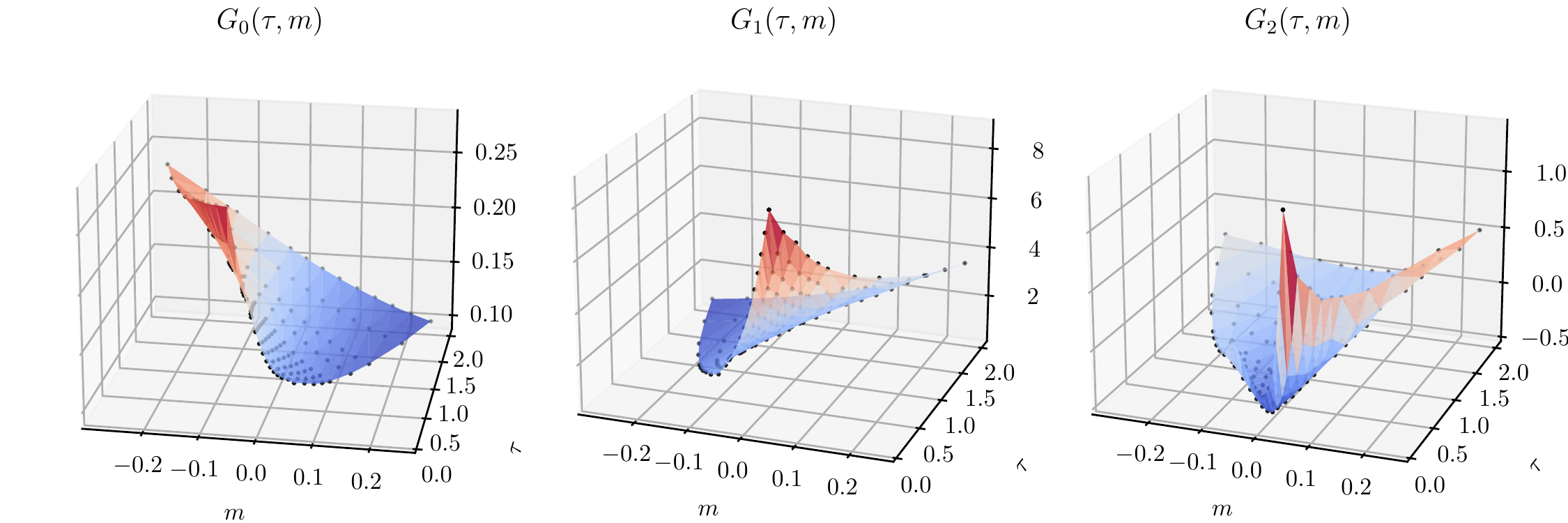}
    \caption{Price basis functions of the normalised call price surface.}
    \label{fig:Gs_primary}
\end{figure}

We plot the price basis functions of these two factors, denoted as $G_1$ and $G_2$, as well as $G_0$, the constant term of $\tilde{c}$, in Figure \ref{fig:Gs_primary}. The points in the liquid lattice (in $(\tau, m)$ coordinates) are also shown on this plot. Here the real-valued functions $G_1$ and $G_2$ are obtained by interpolating the price basis vectors $\mathbf{G}_1$ and $\mathbf{G}_2$, and $G_0$ is obtained by interpolating the normalised call prices averaged over time.

\begin{figure}[!ht]
    \centering
    \includegraphics[scale=.66]{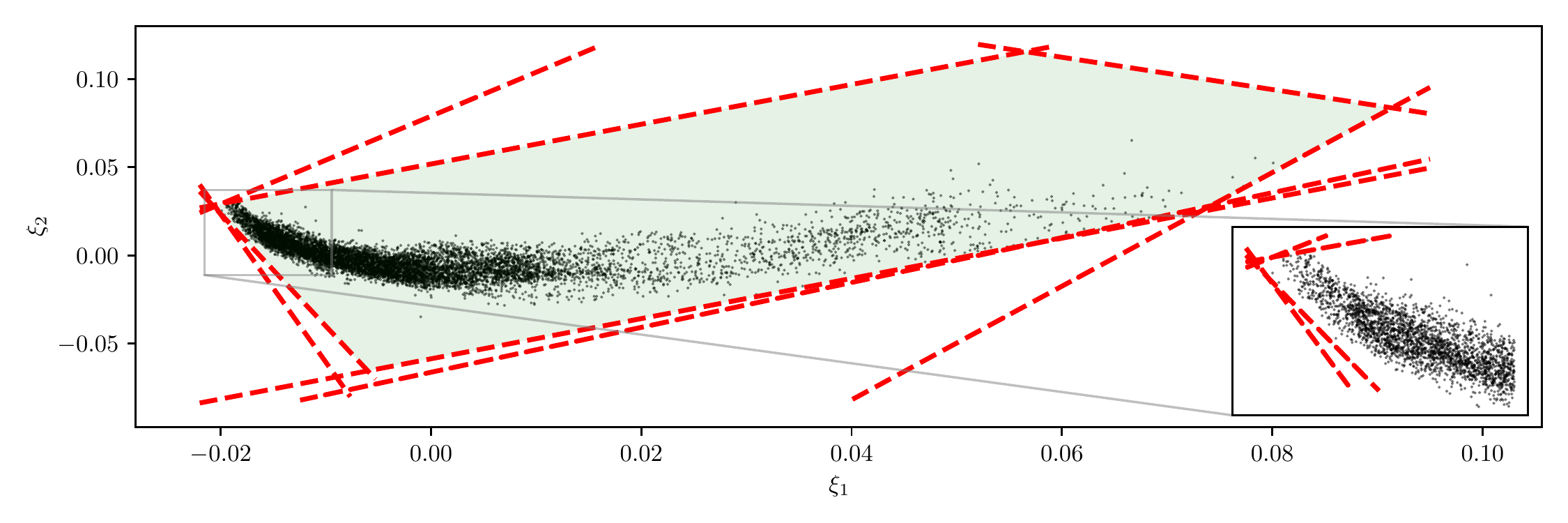}
    \caption{Trajectory (black dots) of the primary $\mathbb{R}^2$ factors and the corresponding static arbitrage constraints (red dashed lines) projected to the $\mathbb{R}^2$ factor space.}
    \label{fig:factors}
\end{figure}

Given these factors, we use the linear programming method (Caron, McDonald and Ponic \cite{caron1989}) to eliminate redundant constraints in the system $\mathbf{A} \mathbf{G}^\top \xi \geq \mathbf{b}$, which is the projection of the original no-arbitrage constraints, constructed in price space, to the $\mathbb{R}^2$ factor space. This results in only 8 constraints, which we indicate as red dashed lines in Figure \ref{fig:factors}. The convex polygonal domain bounded by these constraints (light green area) is the statically arbitrage-free zone for the factors, that is, provided the factor process remains in this region, we are guaranteed to have no static arbitrage in the reconstructed call prices on our liquid lattice $\mathcal{L}_{\text{liq}}$.
Shown in Figure \ref{fig:factors} is also
the cloud of factor realisation decoded from the market data.
We observe that it has a qualitatively similar shape to that
in \cite{cohen2021mktmdl} for simulated prices from a Heston SLV model.

\begin{figure}[!ht]
    \centering
    \includegraphics[scale=.66]{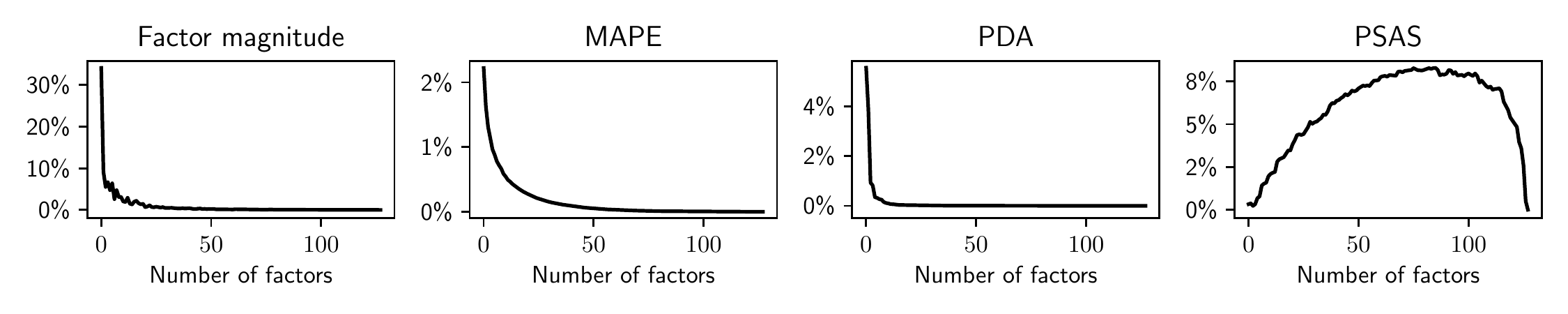}
    \caption{Price reconstruction metrics with the inclusion of more secondary factors.}
    \label{fig:secondary_factor_stats}
\end{figure}

To further reduce the factor reconstruction error, we include a few more secondary factors, whose dynamics are modelled as mean-reverting OU processes. We take the residuals from the reconstruction of call prices with two primary factors, and decode secondary factors consecutively (using PCA) from the residuals, with the objective of maximising statistical accuracy. In Figure \ref{fig:secondary_factor_stats}, we show how various metrics change with the inclusion of more secondary factors. In the left most plot, the factor magnitude is defined as the ratio of the min-max range of the corresponding factor over that of the first primary factor, provided unit-norm price basis vectors. It measures the relative importance of each secondary factor compared to the first primary factor, in reconstructing option prices. While MAPE and PDA decrease with inclusion of more secondary factors, PSAS is not always reduced. As an example, we choose to include 13 secondary factors resulting in MAPE $=0.48\%$, PDA $=0.06\%$ and PSAS $=2.42\%$.

\begin{remark}
PSAS initially increases with more secondary factors included and drops to zero eventually. This is because adding secondary factors leads to small perturbations of the price surface, potentially violating monotonicity and convexity and thus resulting in (small) arbitrages in reconstructed prices. 
In other words, the OU model for the secondary factors does not necessarily rule out static arbitrage, particularly because their drift and diffusion do not satisfy Friedman and Pinsky's condition.
This is less important when choosing the number of secondary factors, firstly as we no longer need to remove arbitrageable data points to estimate their dynamics. 
Moreover, the magnitudes of secondary factors are much smaller than the primary ones (left most plot in Figure \ref{fig:secondary_factor_stats}). In addition, we could always apply the arbitrage repair algorithm \cite{Cohen2020} to perturb our simulated secondary factors in order to restore monotonicity and convexity.
\end{remark}

\subsection{Factor dynamics estimation}

\begin{figure}[!ht]
    \centering
    \includegraphics[scale=.66]{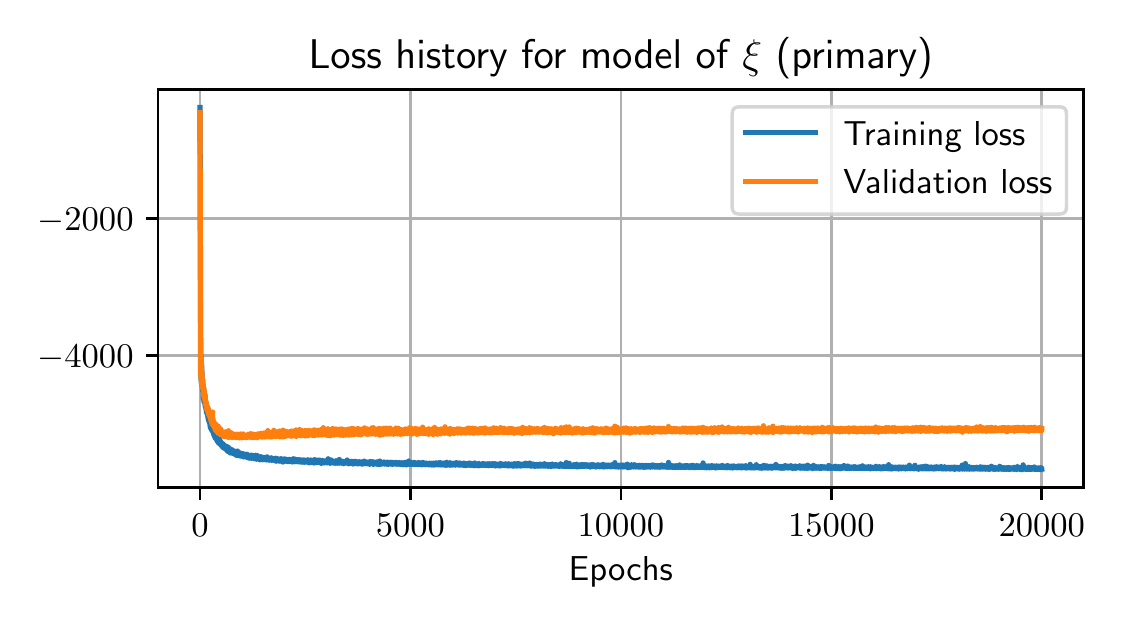}
    \caption{Evolution of training losses and validation losses for the model of the primary factors.}
    \label{fig:loss_hist_xi}
\end{figure}

The estimation of the SDE model for primary factors \eqref{eq:market_model_xi} with static arbitrage constraints has been formulated as an unconstrained deep learning problem in Section 3 of \cite{cohen2021mktmdl}. In this study, we use the network architecture
\begin{equation}
\phi^\theta = \mathcal{F}_{2} \circ \mathcal{A}_\text{ReLU} \circ \mathcal{F}_{256} \circ \mathcal{A}_\text{ReLU} \circ \mathcal{F}_{256} \circ \mathcal{A}_\text{ReLU} \circ \mathcal{F}_{256},
\label{eq:nn_architecture_xi}
\end{equation}
with weight and bias parameters $\theta$,
where $\mathcal{F}_x$ is a fully connected layer, or affine transformation, with $x$ units, and $\mathcal{A}_\text{xxx}$ is an activation function. Each layer $\mathcal{F}$ is parametric, but we omit the parameters for notational simplicity. In addition, to mitigate over-fitting problems, we train both networks with $50\%$ sparsity, meaning that the $50\%$ smallest weights are pruned to zero. We implement and train our model using the standard tools within the Tensorflow \cite{tensorflow2015-whitepaper} environment. In Figure \ref{fig:loss_hist_xi}, we show the evolution of training losses and validation losses\footnote{The first $90\%$ of the training samples are used for training and the last $10\%$ are reserved as validation data.} over epochs during the training of the model. The loss value has a rapid decline for the first a few epochs, and then gradually converges.

\begin{figure}[!ht]
    \centering
    \includegraphics[scale=.66]{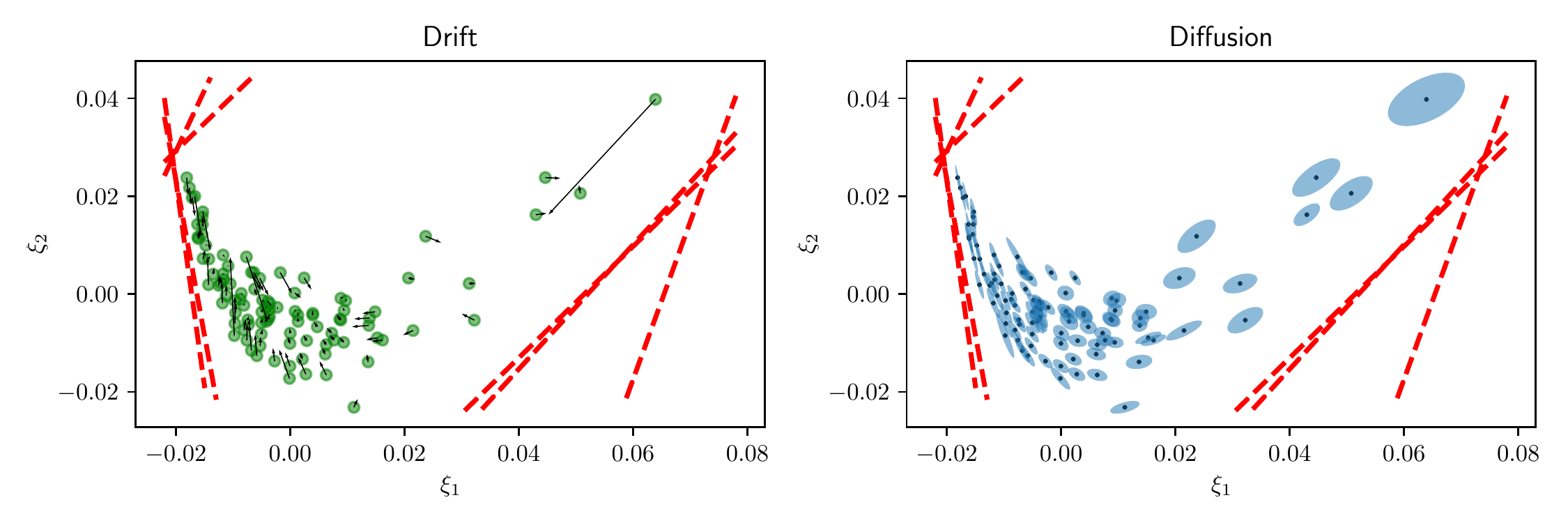}
    \caption{Drift vectors (arrows on the left plot) and diffusion matrices (ellipses representing the principal components of the diffusion on the right plot) for some randomly selected factor data points.}
    \label{fig:learnt_drift_diffusion}
\end{figure}

Next, we sample a few factor data points, and visualise their drift and diffusion coefficients in Figure \ref{fig:learnt_drift_diffusion}. As observed in the trajectory of the factor data in Figure \ref{fig:factors}, the dataset is distributed around a lower-dimensional manifold. For samples on the periphery of the dataset, the learnt drifts tend to point inwards to the manifold, while the principal direction of their diffusions tend to align with the closest boundary. This broadly agrees with the phenomena seen in synthetic data in \cite{cohen2021mktmdl}.

For each of the secondary factors, we discretise the OU model using the Euler--Maruyama scheme, which yields an AR(1) model with normal noises, and then fit with the factor time series data using maximum likelihood estimation.

\subsection{Underlying index dynamics estimation}

\begin{figure}[!ht]
    \centering
    \begin{subfigure}[b]{.49\textwidth}
    \centering
        \includegraphics[scale=.66]{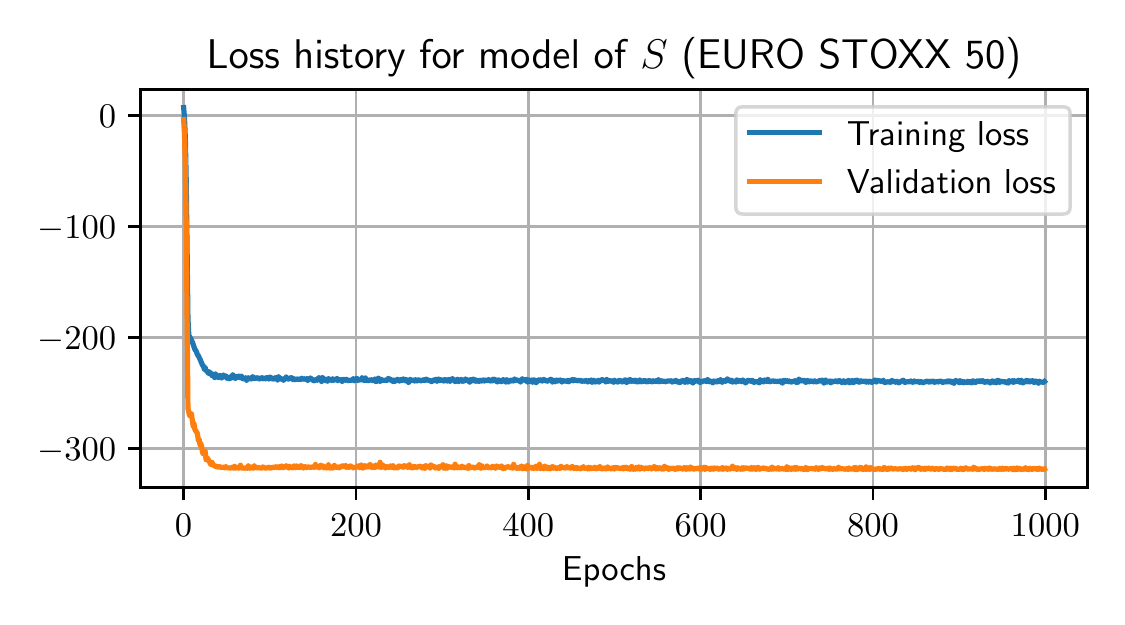}
    \end{subfigure}
    \hfill
    \begin{subfigure}[b]{.49\textwidth}
    \centering
        \includegraphics[scale=.66]{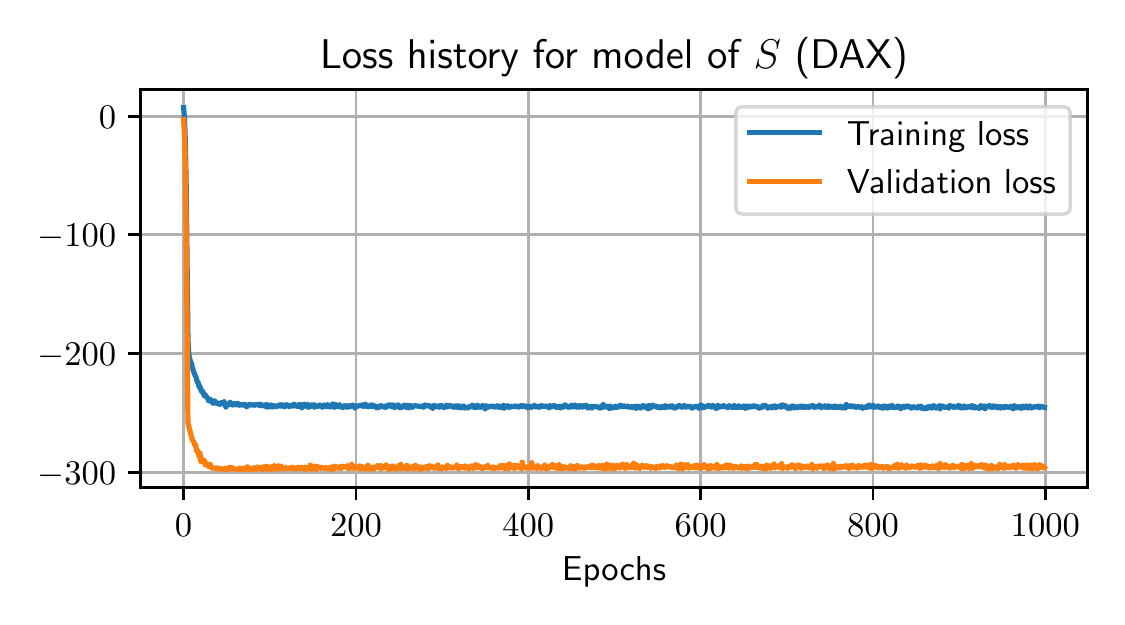}
    \end{subfigure}
    \caption{Evolution of training losses and validation losses for the models of $S$.}
    \label{fig:loss_hist_S}
\end{figure}

Let $X_t = \ln S_t$, then the model for $S$ \eqref{eq:market_model_S} implies, by It\^{o}'s lemma,
\begin{equation*}
    \diff X_t = r(\xi_t) \diff t + \gamma(\xi_t) \diff W_{0,t}, \text{ where } r(\xi_t) = \alpha(\xi_t) - \frac{1}{2} \gamma^2 (\xi_t).
\end{equation*}
Since drift estimation is well known to be noisy, we assume a constant drift for $X_t$, i.e, $r(\xi_t) \equiv r$. With this assumption, an unbiased estimate for $r$ is
\begin{equation}
    \hat{r} = \frac{1}{\delta t} \left( \frac{1}{L-1} \sum_{l=1}^{L-1} \ln \left( \frac{S_{t_{l+1}}}{S_{t_l}} \right) \right) = \frac{1}{T} \ln \left(\frac{S_T}{S_0} \right). 
    \label{eq:S_drift_est}
\end{equation}
Hence, for each underlying equity index, we first estimate a constant drift using historical log-returns, and then train a neural-SDE model with only its diffusion represented by a neural network. In particular, for our training samples (from 2002-01-01 to 2018-12-31), $\hat{r}^U = -0.027$, $\hat{r}^D = 0.071$. Compared with the model for the primary factors, we use a smaller network $\phi^{S,\theta}: \mathbb{R}^2 \rightarrow \mathbb{R}$ for the neural-SDE model of $S$:
\begin{equation}
    \phi^{S,\theta} = \mathcal{F}_{2} \circ \mathcal{A}_\text{ReLU} \circ \mathcal{F}_{128} \circ \mathcal{A}_\text{ReLU} \circ \mathcal{F}_{128} \circ \mathcal{A}_\text{ReLU} \circ \mathcal{F}_{128}.
    \label{eq:nn_architecture_S}
\end{equation}
In Figure \ref{fig:loss_hist_S}, we show the evolution of training losses and validation losses over epochs during the training of the models for both EURO STOXX 50 index and DAX index. The loss value quickly drops during the first 10 epochs, and slowly declines and converges within 50 epochs.

\begin{figure}[!ht]
    \centering
    \begin{subfigure}[b]{.49\textwidth}
    \centering
        \includegraphics[scale=.64]{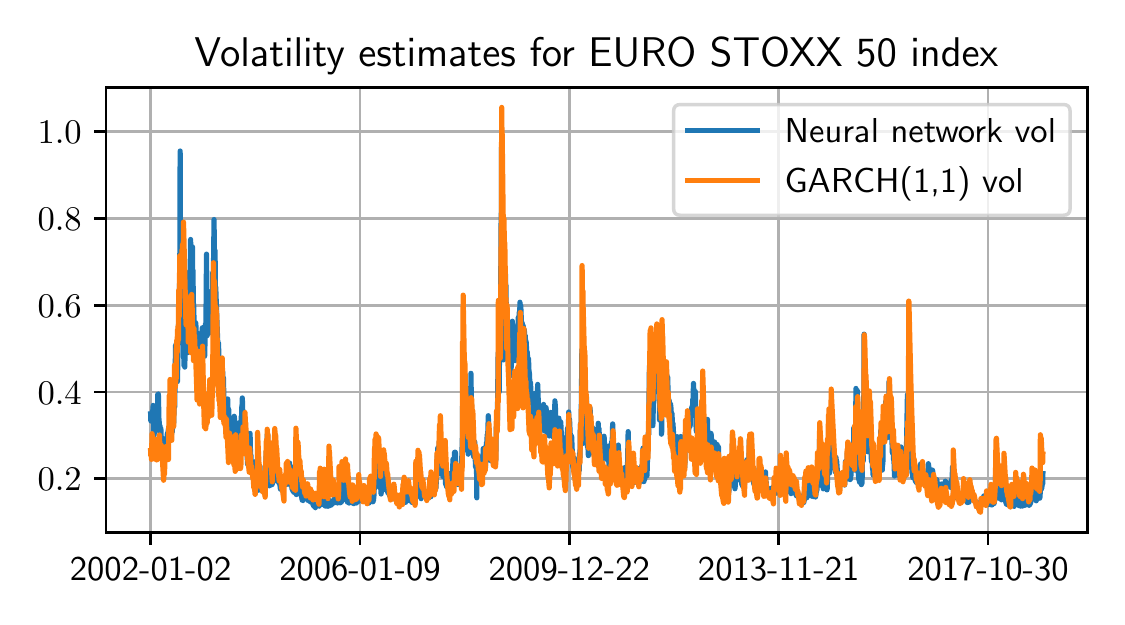}
    \end{subfigure}
    \hfill
    \begin{subfigure}[b]{.49\textwidth}
    \centering
        \includegraphics[scale=.64]{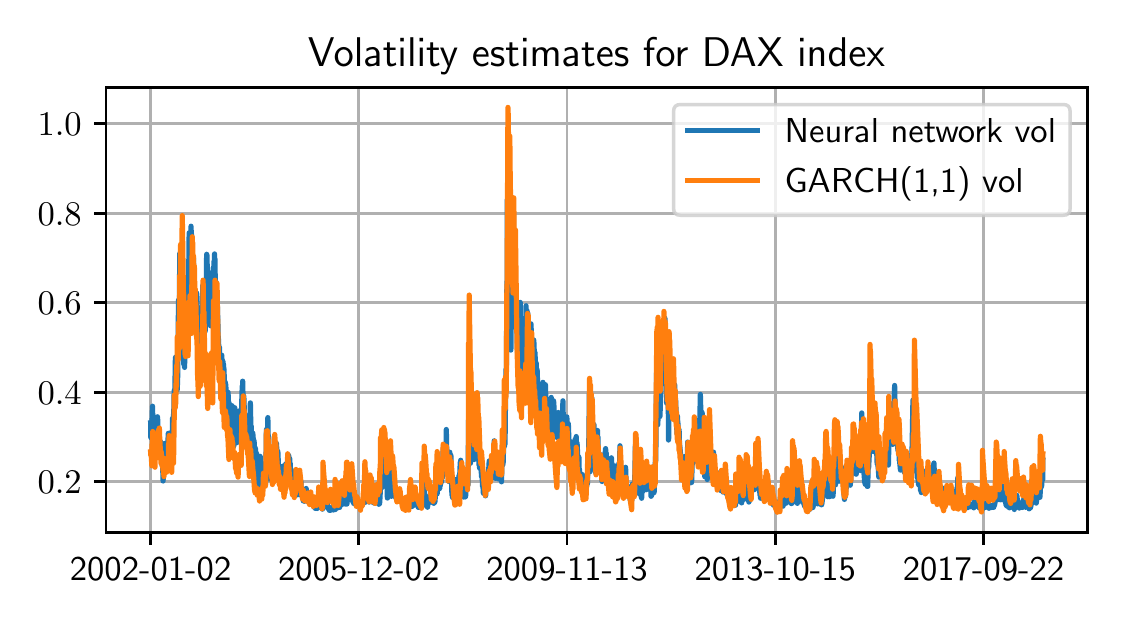}
    \end{subfigure}
    \caption{Comparison of the volatility estimates for the option underlying indices.}
    \label{fig:vol_S}
\end{figure}

To examine how well the neural network models the diffusion of $S$, we use a GARCH(1,1) model to estimate empirical volatilities and compare with the in-sample volatilties computed from the trained neural networks. As evidenced in Figure \ref{fig:vol_S}, the neural network models produce similar volatility dynamics.

\subsection{Analysis of historical residuals}

\begin{figure}[!ht]
    \centering
    \includegraphics[scale=.66]{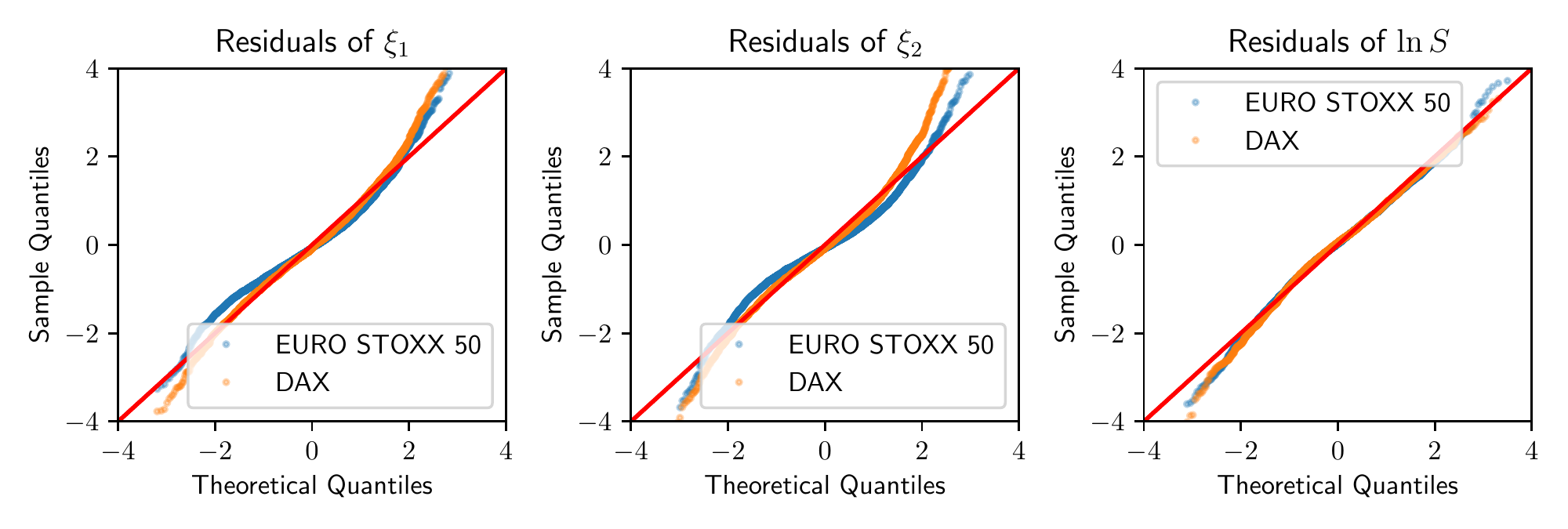}
    \caption{QQ-plots of historical in-sample model residuals.}
    \label{fig:hist_res_marginaldensity}
\end{figure}

Historical residuals of the trained neural-SDE models \eqref{eq:market_model} at time $t_l$, for $l=1,\dots,L-1$, are computed by
\begin{equation*}
    \hat{Z}_{t_l}^\xi = \sigma^{-1}(\xi_{t_l}) \left( \xi_{t_{l+1}} - \xi_{t_{l}} - \mu(\xi_{t_{l}}) \delta t \right), \quad
    \hat{Z}_{t_l}^S = \gamma^{-1}(\xi_{t_l}) \left( \ln \frac{S_{t_{l+1}}}{S_{t_l}} - \hat{r} \delta t \right).
\end{equation*}
Our approximate likelihood corresponds to an assumption that $(Z^S_t, Z^\xi_t)$ are independent standard normal white noise. Therefore, it is worth checking if the estimated residuals $(\hat{Z}^S_t, \hat{Z}^\xi_t)$ appear to be independent standard normals, within our training data. In Figure \ref{fig:hist_res_marginaldensity}, we show the normal QQ-plots of the empirical marginal distributions for the residuals. For the primary factors, their residuals show mild fat tails. For the underlying stock indices, their residuals appear slightly left-skewed, implying higher-than-expected probabilities for large loss.

\begin{figure}[!ht]
    \centering
    \includegraphics[scale=.66]{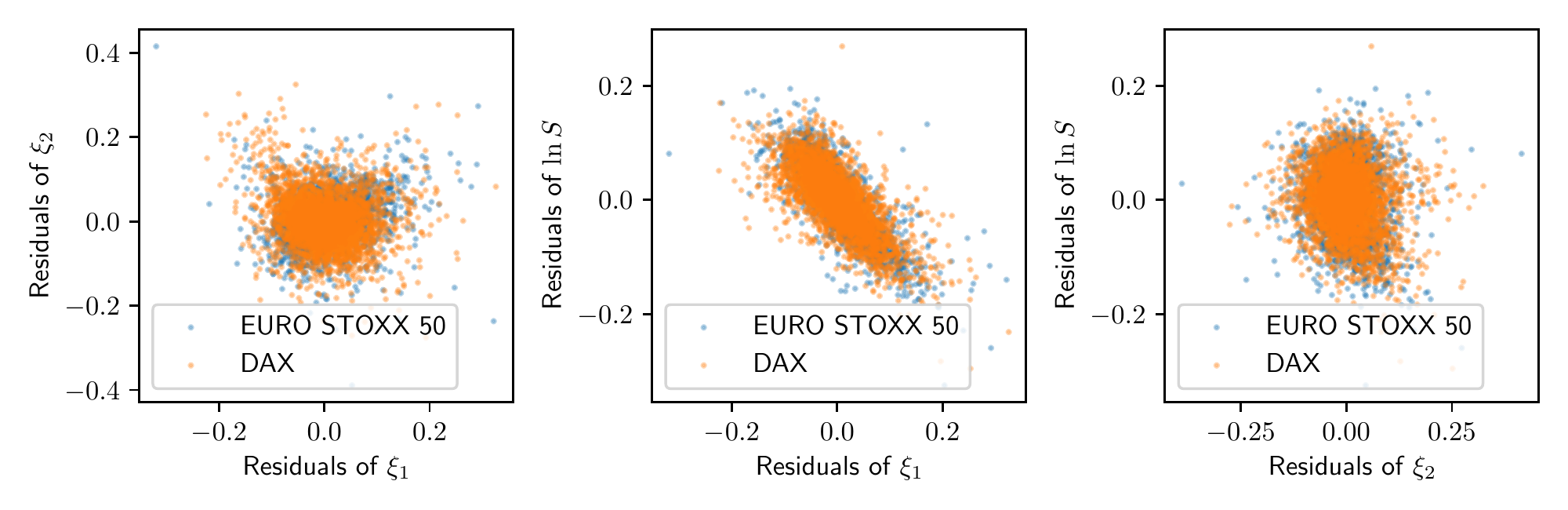}
    \caption{Scattter plots of historical in-sample model residuals.}
    \label{fig:hist_res_jointdensity}
\end{figure}

To investigate dependence between the historical residuals for each factor, we show in Figure \ref{fig:hist_res_jointdensity} the scatter plots of each pair of residuals. The residuals among risk factors $\xi_i$ look fairly uncorrelated, indicating that the trained neural nets have well captured factor covariances over time. However, there is evident negative correlation between the residuals of $\ln S$ and $\xi_1$, coinciding with the commonly seen \textit{leverage effect} \cite{black1976studies}. In fact, we could capture the leverage effect using the neural-SDE model where a full covariance matrix between $(\ln S, \xi)$ is specified, which consequently produces uncorrelated residuals --- see Appendix \ref{apd:model_xiS}.

\subsection{Out-of-sample simulation of factors and implied volatility surface}
\label{sec:factor_simulation}

We assess the learnt model's ability to simulate time series data that are like the input real data. We take the trained model and simulate sample paths using a tamed\footnote{We include the taming method simply to ensure stability of simulations if our neural networks were to produce unusually large values for drifts or volatilities.} Euler scheme (see Hutzenthaler, Jentzen and Kloeden \cite{Hutzenthaler2012}), 
which is given by
\begin{equation}
    \xi_{t + \delta t} = \xi_t + \frac{\mu (\xi_t)}{1 + |\mu(\xi_t)| \sqrt{\delta t}} \delta t + \frac{\sigma(\xi_t)}{1 + \|\sigma(\xi_t)\| \sqrt{\delta t}} Z^\xi_t, 
    ~\ln \frac{S_{t+\delta t}}{S_t} = \hat{r} \delta t + \frac{\gamma(\xi_t)}{1 + |\gamma(\xi_t)| \sqrt{\delta t}} Z^S_t,
\label{eq:tamed_euler}
\end{equation}
where the values of $\mu$, $\sigma$ and $\gamma$ are approximated by the trained neural networks. Rather than generating the innovations $(Z^S_t, Z^\xi_t)$ from normal distributions, we randomly sample them from historical residuals of the trained models. There are two major advantages of drawing innovations from historical residuals rather than normal distributions. First, historical residuals have fatter tails than normal (see Figure \ref{fig:hist_res_marginaldensity}). Second, the joint historical residuals implicitly preserve the (potentially higher-order) dependence of $(S, \xi)$ that fails to be captured by the covariance terms (see Figure \ref{fig:hist_res_jointdensity}).

\begin{figure}[!ht]
    \centering
    \includegraphics[scale=.66]{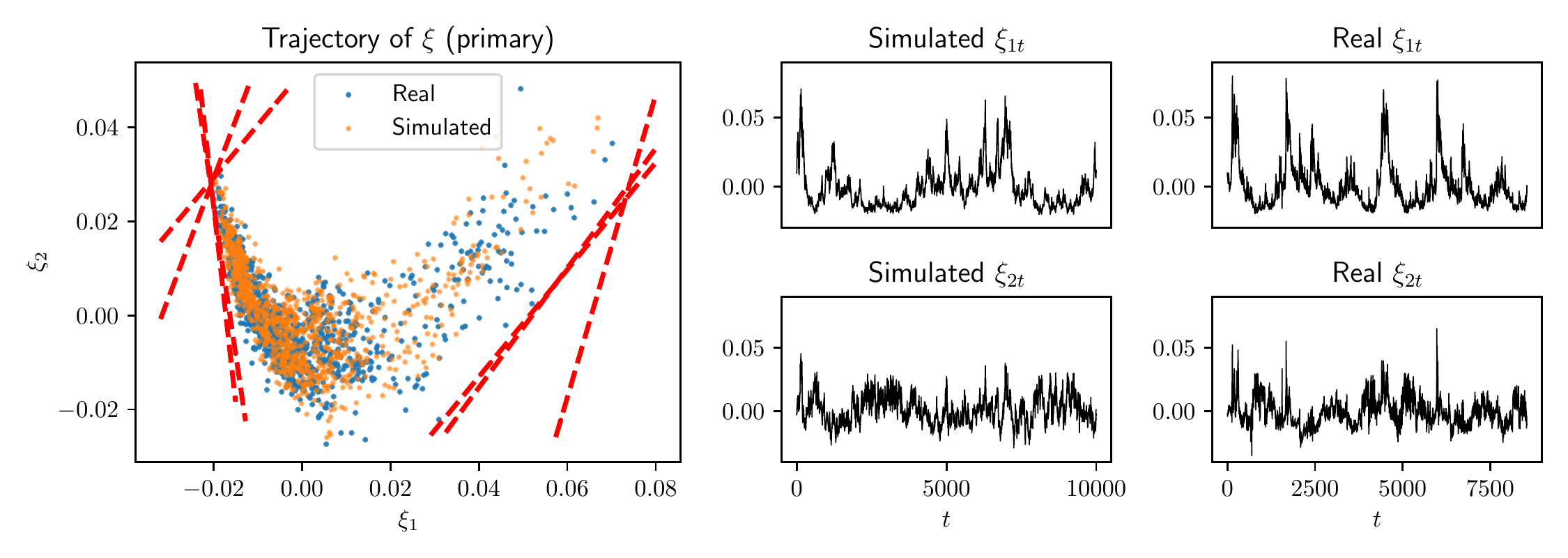}
    \caption{Simulation of primary factors from the learnt neural-SDE model, compared with the real data.}
    \label{fig:sim_fators}
\end{figure}

\begin{figure}[!ht]
    \centering
    \includegraphics[scale=.66]{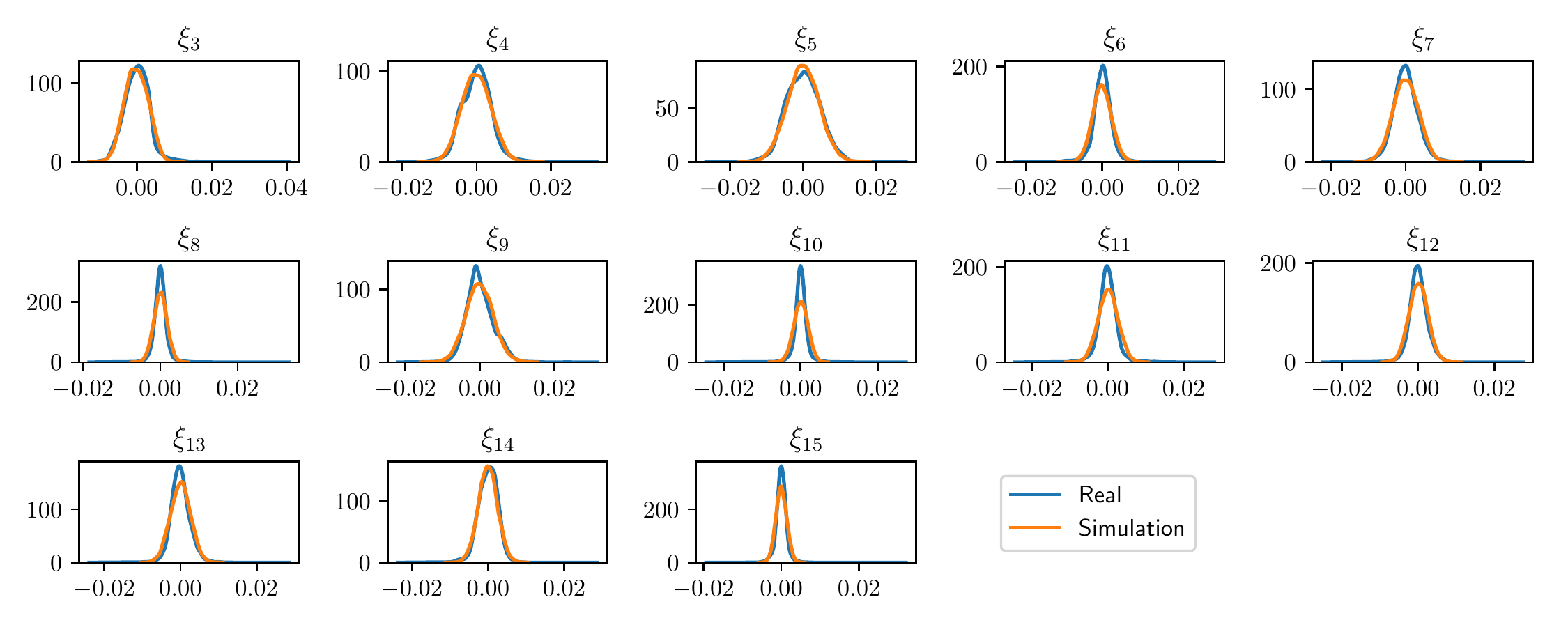}
    \caption{Distribution of OU-simulated and real secondary factors.}
    \label{fig:sim_factor_dist_sec}
\end{figure}

We show the time series of $\xi$ in Figure \ref{fig:sim_fators} (right). In the scatter plot on the left of Figure \ref{fig:sim_fators}, we see that the dependence structure between $\xi_1$ and $\xi_2$ is well captured. In addition, the simulated factors remain within the no-arbitrage region, due to the hard constraints imposed on the drift and diffusion functions. We also simulate secondary factors from their fitted OU processes, and check the similarity in distribution, as shown in Figure \ref{fig:sim_factor_dist_sec}.

\begin{figure}[!ht]
    \centering
    \begin{subfigure}[b]{.24\textwidth}
    \centering
        \includegraphics[scale=.66]{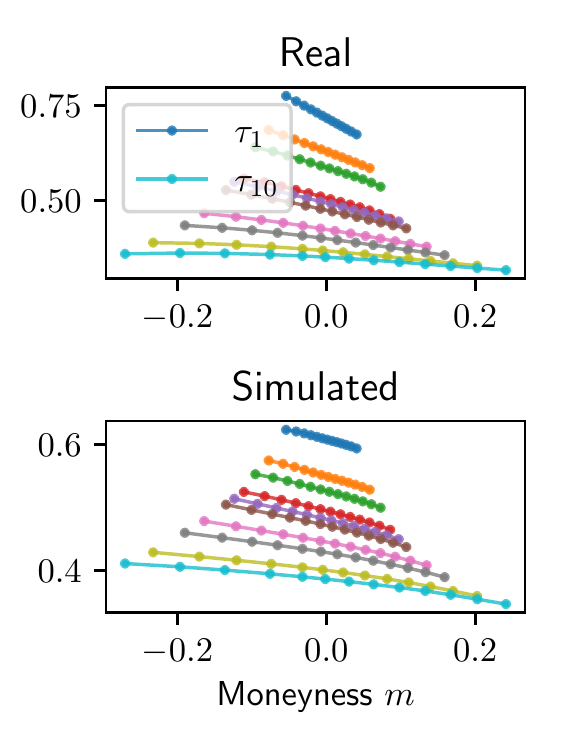}
        \caption{ES50 2008-10-17}
    \end{subfigure}
    \hfill
    \begin{subfigure}[b]{.24\textwidth}
    \centering
        \includegraphics[scale=.66]{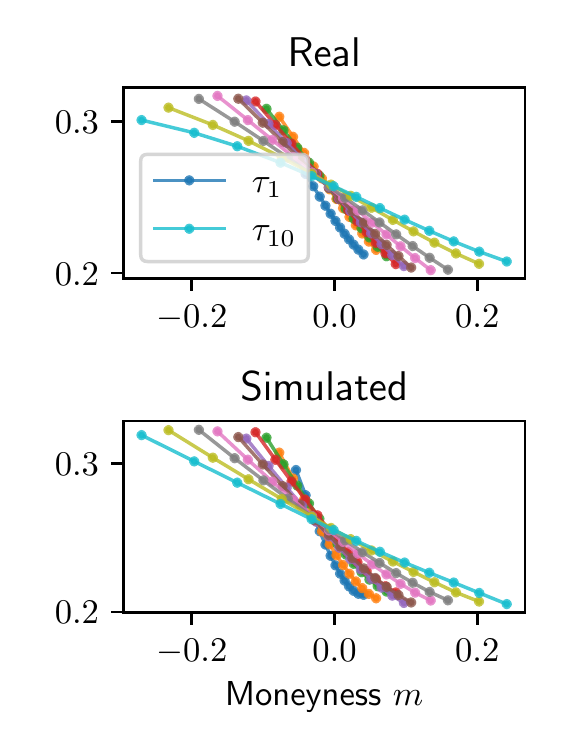}
        \caption{ES50 2010-07-29}
    \end{subfigure}
    \hfill
    \begin{subfigure}[b]{.24\textwidth}
    \centering
        \includegraphics[scale=.66]{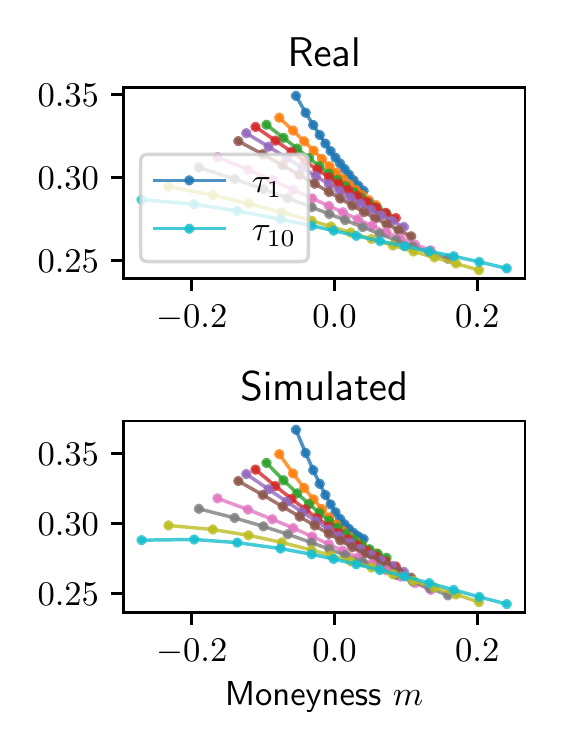}
        \caption{DAX 2002-05-29}
    \end{subfigure}
    \hfill
    \begin{subfigure}[b]{.24\textwidth}
    \centering
        \includegraphics[scale=.66]{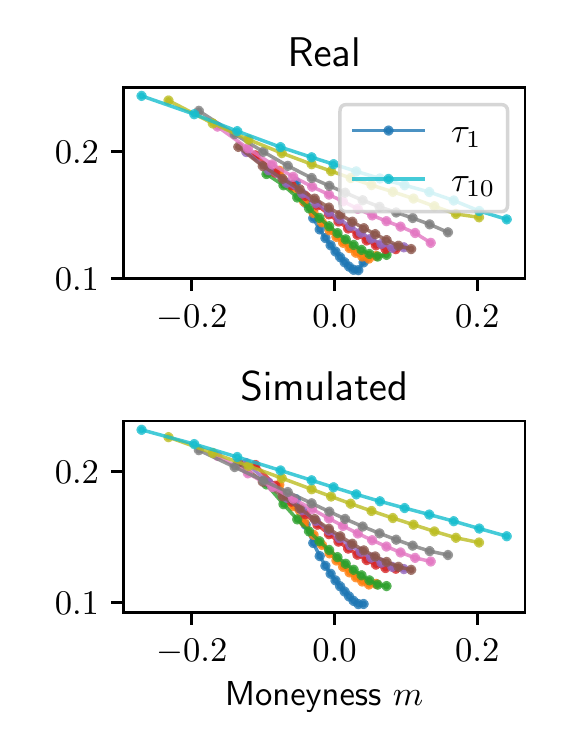}
        \caption{DAX 2017-07-18}
    \end{subfigure}
    \caption{Examples of real and simulated implied volatility surfaces.$\tau_1=30$ days and $\tau_{10}=730$ days.}
    \label{fig:iv_sim}
\end{figure}

The learnt model is also capable of simulating a variety of realistic patterns of implied volatility surface (IVS). As demonstrated in Figure \ref{fig:iv_sim}, we pick a few IVS patterns observed in historical data and find the closest (in the sense of minimising the $\ell^2$ distance for $\xi$) ones in the simulated sample path.

\subsection{VIX-like volatility index simulation}
\label{ref:vix}

We follow the CBOE VIX calculation methodology \cite{vix} to compute a \textit{volatility index} from the real and simulated option prices. Specifically, VIX is a linear combination of OTM call and put option prices, which can be further written as a linear combination of call prices only, provided that put-call parity holds under no-arbitrage.

\begin{figure}[!ht]
    \centering
    \begin{subfigure}[b]{.39\textwidth}
    \centering
        \includegraphics[scale=.66]{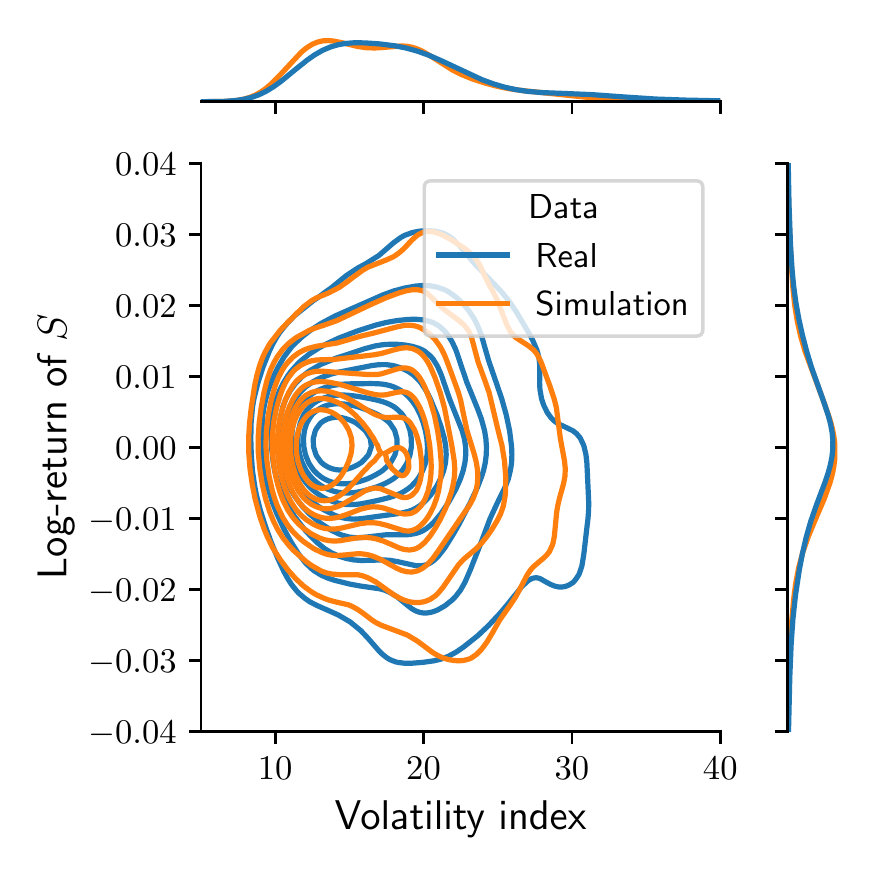}
    \end{subfigure}
    \hfill
    \begin{subfigure}[b]{.59\textwidth}
    \centering
        \includegraphics[scale=.66]{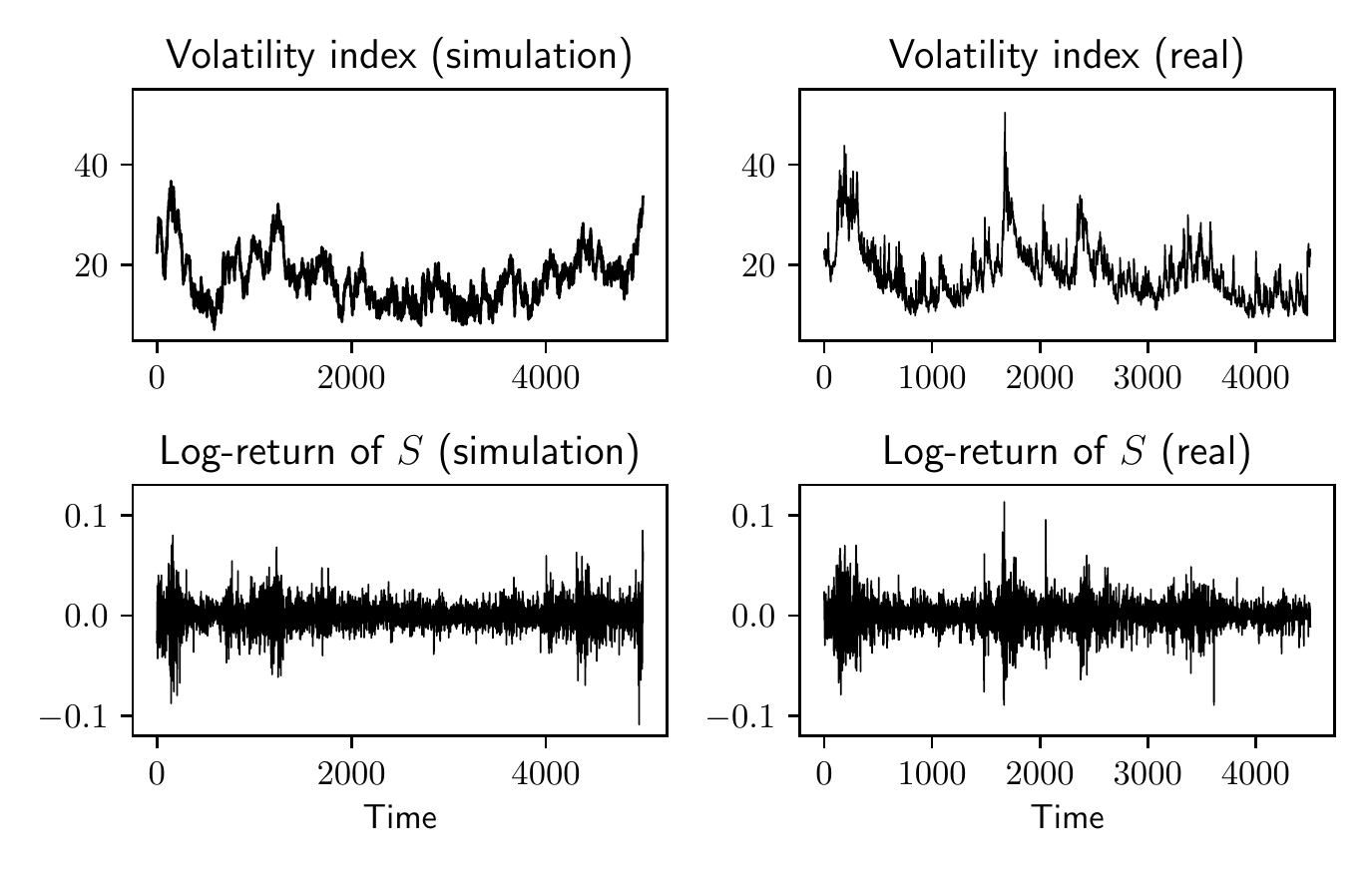}
    \end{subfigure}
    \caption{Simulation of the VIX-like volatility index and log-return of $S$ (EURO STOXX 50 index), compared with the real data.}
    \label{fig:EX50_S_vix}
\end{figure}

In Figure \ref{fig:EX50_S_vix}, we compare the real historical distribution of (VIX, log-return of $S$) and one simulation using the trained market model. For the simulated VIX, both its marginal distribution and the joint distribution with the log-return of $S$ look reasonably similar to those of the real data. This demonstrates that our model is capturing the dependence structure between the volatility index and the underlying $S$. Looking at the simulated time series of VIX and log-return of $S$, we see several occurrences of volatility clustering in the return series, which always coincide with high VIX values.

\section{Risk management}
\label{sec:risk_mgt}

As our market models are defined and trained under the real-world measure, a natural application is to simulate risk scenarios for option portfolios. We explain how this is done, and compare its performance with the filtered historical simulation method \cite{fhs2015}, which is a standard approach used widely for managing risk of futures and options by mainstream CCPs (see, for example, CME \cite{cme_im}, Eurex \cite{eurex_im} and LCH \cite{lch_im}).

\subsection{Measuring risk}

Consider a portfolio $\Pi$ of $N$ call options at time $t$ with market value $\Pi_t = \sum_{j=1}^N w_j C_t(T_j, K_j)$, where $w_j \neq 0$ denotes the weight of the $j$-th call option in the portfolio. The profit-and-loss (PnL) of the portfolio at time $t + \Delta t$, for given time horizon $\Delta t$, is then
\begin{equation*}
    \text{PnL}_{\Delta t} (\Pi_t) = \Pi_{t+\Delta t} - \Pi_{t} = \sum_{j=1}^N w_j \left( C_{t+\Delta t} (T_j, K_j) - C_{t} (T_j, K_j) \right) = \sum_{j=1}^N w_j \text{PnL}_{\Delta t} (C_t(T_j, K_j)).
\end{equation*}
For a chosen risk measure $\varrho(\cdot)$, time $t$ and horizon $\Delta t$, we evaluate the risk of the portfolio, $\varrho (\text{PnL}_{\Delta t} (\Pi_t))$,
in broadly three steps:
\begin{enumerate}[label=(\arabic*)]
    \item Identify \textit{risk factors} $\Theta$ and a \textit{value function} $V$ such that $\Pi_t \approx V(t, \Theta_t)$.
    \item Simulate $\Theta_{t+\Delta t}^{(m)} | \Theta_t$ based on some dynamic model for $\Theta_t$ for $m=1,\dots, M$, and compute $\Pi_{t+\Delta t}^{(m)} \approx V(t+\Delta t, \Theta_{t+\Delta t}^{(m)})$. Each $\Pi_{t+\Delta t}^{(m)}$ is called a \textit{risk scenario}.
    \item For the $m$-th risk scenario, calculate the \textit{scenario PnL} as $V(t+\Delta t, \Theta_{t+\Delta t}^{(m)}) - \Pi_t$, which is assumed to be a realisation of $\text{PnL}_{\Delta t} (\Pi_t) | \Theta_t$. Therefore, the $M$ scenario PnLs give an estimate of the empirical distribution\footnote{In general, the conditional distribution of PnL is not available in closed-form, otherwise one may be able to derive an analytical formula for the risk measure without simulation.} of $\text{PnL}_{\Delta t} (\Pi_t) | \Theta_t$, which the risk measure $\varrho$ will be evaluated on.
\end{enumerate}

\subsection{Risk simulation using the market model}

Our market models characterise option risk in terms of $\Theta = (S, \xi)$, for which the joint dynamics are given as the SDE system \eqref{eq:market_model}. After an offline training of the SDE model, we discretise it and simulate the factors, which are subsequently translated to arbitrage-free option prices through simple linear transformations.

More specifically, suppose we want to simulate prices for the $(T,K)$-call option at time $t+\Delta t$. Then the simulation consists of the following steps:
\begin{enumerate}[label=(\arabic*)]
    \item \textbf{Decode factors at time $t$.}
    We collect prices of call options on the liquid lattice $\mathcal{L}_\text{liq}$ and normalise the prices using the time-$t$ zero curve and futures curve. Given the price basis $\mathbf{G}^\top$, we solve for $\xi_t = (\mathbf{G} \mathbf{G}^\top)^{-1} \mathbf{G} (\mathbf{c}_t - \mathbf{G}_0^\top)$.
    
    \item \textbf{Simulate factors at time $t + \Delta t$.}
    Given the SDE model \eqref{eq:market_model} for which the coefficients are estimated by neural networks, we simulate $S^{(m)}_{t+\Delta t}$ and $\xi^{(m)}_{t + \Delta t}$ using the tamed Euler scheme \eqref{eq:tamed_euler} iteratively over $\Delta t / \delta t$ time steps, for $m=1,\dots,M$.
    
    \item \textbf{Compute option prices at time $t+\Delta t$.}
    Under the $m$-th simulated scenario, we compute the normalised prices for the call options on $\mathcal{L}_\text{liq}$ as $\mathbf{c}_{t + \Delta t}^{(m)} = \mathbf{G}^\top_0 + \mathbf{G}^\top \xi_{t+\Delta t}^{(m)}$. Let $h(\tau, m; \mathbf{c})$ be some $C^{1,2}$ function that interpolates the data $(\mathcal{L}_\text{liq}, \mathbf{c})$, then
    \begin{equation*}
        C_{t+\Delta t}^{(m)}(T,K) = S^{(m)}_{t+\Delta t} \times h \left(\tau_t - \Delta t, m_t + \ln \frac{S_t}{S^{(m)}_{t+\Delta t}} ;\mathbf{c}_{t + \Delta t}^{(m)}\right),
    \end{equation*}
    where $\tau_t = T - t$ and $m_t = \ln(K/F_t{(T)})$.
\end{enumerate}

\begin{remark}
Our model is written not in terms of prices $C$ but in terms of the normalised prices $\tilde{c}$. This transformation \eqref{eq:call_price_transformation} involves discounting both for interest rates and dividends. While interest rates and dividends could be included as risk factors, they have only marginal impact on equity index options, especially when the longest expiry we consider is two years. We will therefore restrict our attention to the underlying risk and volatility risk. Therefore, we will evaluate risk of the revised option value $C_t = S_t \tilde{c}_t$.
\end{remark}

\subsection{Filtered historical simulation with Heston models}%

The historical simulation approach \cite{barone1997var} simply uses historical changes of a collection of risk factors $\Theta = (x_1, \dots, x_d)$ to construct risk scenarios. In addition, historical changes of risk factors may be \textit{filtered} (or \textit{scaled}) based on current volatility, so that the simulated risk scenarios are more responsive to recent market conditions. Specifically, suppose $C_t(T,K) = V(T-t, K; \Theta_t)$ and we want to simulate prices at $t+\Delta t$. Then there are two steps:
\begin{enumerate}[label=(\arabic*)]
    \item Collect historical data of $\Delta t$-returns $r_j$ in the $j$-th risk factor $x_j$, for $j=1,\dots, d$, over the last $M$ timesteps, so we have $r_{j}(t_m)$ for $m=1,\dots,M$. Let $\sigma_j(t)$ be some realised volatility estimate for the $j$-th risk factor at $t$. Then we define
    \begin{equation}
        r^{(m)}_j(t+\Delta t) = r_j(t_m) \frac{\sigma_j(t)}{\sigma_j(t_m)}
        \label{eq:filtered_return}
    \end{equation}
    as the risk factor return under the $m$-th scenario, which leads to the corresponding risk factor scenario $x_j^{(m)}(t+\Delta t)$ by applying the return\footnote{We use log returns if a risk factor is positive by definition, in which case $x_j^{(m)}(t+\Delta t) = x_j(t) \exp({ r^{(m)}_j(t+\Delta t)})$, or absolute return otherwise, in which case $x_j^{(m)}(t+\Delta t) = x_j(t) + { r^{(m)}_j(t+\Delta t)}$.} to $x_j(t)$.
    \item Revalue the call option $C_{t+\Delta t}^{(m)} (T,K) = V(T-t-\Delta t, K; \Theta_{t+\Delta t}^{(m)})$.
\end{enumerate}

We use the parameters of calibrated Heston models \cite{Heston1993} as risk factors. Specifically, we calibrate the Heston parameters $\Theta = (S_0, \nu_0, \theta, k, \sigma, \rho)$ as detailed in Appendix \ref{apd:heston_calib} to historical vega-weighted option prices, yielding historical observations of $\Theta_t$, which are then used to simulate historical scenarios for option portfolios.

\subsection{Value-at-Risk backtesting analysis}
\label{sec:bar_backtesting}

In order to assess the scenarios produced by risk simulation engines for a comprehensive set of risk profiles, we backtest various option portfolios ranging from outright options to heavily hedged portfolios. The list of tested portfolios is given in Table \ref{tab:test_portfolio}, with detailed description in Appendix \ref{apd:test_strategy}. We consider both long and short positions in each portfolio considered.

\begin{table}[!ht]
    \centering
    \footnotesize
\begin{tabular}{ccccccccc}
\toprule
\multicolumn{1}{l}{} & \multicolumn{3}{c}{Delta-exposed} & \multicolumn{5}{c}{Delta-hedged} \\ \cmidrule(lr){2-4} \cmidrule(lr){5-9}
\textbf{Portfolio}            & Outright & \begin{tabular}[c]{@{}c@{}}Delta\\ spread\end{tabular} & \begin{tabular}[c]{@{}c@{}}Risk\\ reversal\end{tabular} & \begin{tabular}[c]{@{}c@{}}Delta\\ butterfly\end{tabular} & \begin{tabular}[c]{@{}c@{}}Delta-hedged\\ option\end{tabular} & \begin{tabular}[c]{@{}c@{}}Delta-neutral\\ strangle\end{tabular} & \begin{tabular}[c]{@{}c@{}}Calendar\\ spread\end{tabular} & VIX \\
\cmidrule(lr){1-1} \cmidrule(lr){2-4} \cmidrule(lr){5-9}
\textbf{Number} & 140 & 420 & 60 & 20 & 60 & 60 & 90 & 2 \\
\bottomrule
\end{tabular}
    \caption{Number of tested portfolios of various types.}
    \label{tab:test_portfolio}
\end{table}

We have split the historical data into training and testing samples according to Table \ref{tab:data}. We use only the training samples for estimating the neural-SDE market model, which will then be used to calculate risks. As a measure of risk of a portfolio $\Pi$, we consider Value-at-Risk (VaR) at a given confidence level $\alpha$ for a given time horizon $\Delta t$, defined as
\begin{equation*}
    \text{VaR}_{\alpha, \Delta t} (\Pi_t) = - \sup \left\{l \in \mathbb{R}: \mathbb{P} \left( \text{PnL}_{\Delta t} (\Pi_t) < l \right) \leq 1-\alpha \right\}.
\end{equation*}
Given $M$ risk simulation scenarios, the VaR is estimated by
\begin{equation*}
    \widehat{\text{VaR}}_{\alpha, \Delta t} (\Pi_t) = - \sup \left\{l \in \mathbb{R}: \frac{1}{M} \sum_{m=1}^M \mathbbm{1}_{\left\{ \text{PnL}^{(m)}_{\Delta t} (\Pi_t) < l \right\}} \leq 1-\alpha \right\}.
\end{equation*}

For each testing sample $t$, we compute an ex-ante VaR forecast $\widehat{\text{VaR}}_{\alpha, \Delta t} (\Pi_t)$, and observe an ex-post realised $\text{PnL}_{\Delta t} (\Pi_t)$. Therefore, we can define the VaR breach
\begin{equation*}
\beta(\Pi_t) = \mathbbm{1}_{\{ \text{PnL}_{\Delta t} (\Pi_t) < -\widehat{\text{VaR}}_{\alpha, \Delta t} (\Pi_t) \}}.
\end{equation*}
The coverage ratio over the $L$ testing periods is then computed as
\begin{equation*}
    1 - \frac{1}{L-E} \sum_{l=E+1}^L \beta(\Pi_{t_l}).
\end{equation*}

Prevailing VaR model backtesting methodologies mostly focus on two categories:
\begin{enumerate}[label=(\arabic*)]
    \item \textbf{Coverage tests} assess whether the empirical frequency of VaR breaches is consistent with the quantile of loss a VaR measure is intended to reflect. We use the widely cited Kupiec's proportion of failures (PF) coverage test  \cite{kupiec1995}, which assumes the null $H_0: \mathbb{E}[\beta(\Pi_t) | \Theta_t] = 1-\alpha$ and models the VaR breach events in each testing sample as independent Bernoulli trials. In addition, we also consider the Basel Committee traffic light approach \cite{basel1996}: based on the number of breaches experienced for the 1-day $\text{VaR}_{0.99}$ results during a 250-day testing window, the VaR measure is categorised as falling into the green ($\leq 4$), yellow ($\geq 5$ and $\leq 9$) or red ($\geq 10$) zone.
    
    \item \textbf{Independence tests} assess whether VaR breach events appear to be independent of each other. We use Christoffersen's test \cite{christoffersen1998} that considers unusually frequent consecutive breaches. Combining the test statistics of the Kupiec's PF test and the Christoffersen's independence test yields a conditional coverage (CC) mixed test \cite{christoffersen1998}.
\end{enumerate}

We compute VaR for horizons $\Delta t = 1, 2, 5$ and $10$ days, where 2-day and 5-day are typical \textit{margin periods of risk} that European regulators require for risk-monitoring of exchange-traded and over-the-counter derivatives, respectively.

We report the backtesting coverage and independence statistics for 1-day VaRs of EURO STOXX 50 index options at 0.99 and 0.95 confidence levels in Table \ref{tab:backtesting_mpor1}. We compute a coverage ratio for each of the tested option portfolios, and calculate its sample median and mean over the portfolios. The neural-SDE VaR model gives a higher mean coverage ratio over option portfolios, while much fewer portfolios fail the coverage and independence tests. At the 0.95 confidence level, more portfolios fail the two-sided Kupiec PF coverage test than the one-sided version, indicating that both VaR models tend to overestimate risks, but the neural-SDE approach is less prone to underestimating risks (as indicated by the one-sided test failure rate of 9.39\% v.s. 18.08\%)\footnote{It is worth noting that since we reject the nulls in all the coverage and independence tests at the 0.05 significance level, a perfect VaR model is expected to fail for 5\% of portfolios.}.

\begin{table}[!ht]
\footnotesize
\centering
\begin{tabular}{lcccc}
\toprule
\multirow{2}{*}{}                             & \multicolumn{2}{c}{1-day $\text{VaR}_{0.99}$} & \multicolumn{2}{c}{1-day $\text{VaR}_{0.95}$} \\ \cmidrule(lr){2-3} \cmidrule(lr){4-5}
                            & nSDE &  FHS & nSDE  & FHS \\ \cmidrule(lr){1-5}
Coverage ratio median       & 0.9921 & 0.9881 & 0.9484 & 0.9563 \\
Coverage ratio mean         & 0.9887 & 0.9742 & 0.9528 & 0.9321 \\
Kupiec PF (two-sided)      & 6.92\% & 25.23\% & 32.51\% & 33.69\% \\
Kupiec PF (one-sided)      & 6.92\% & 25.23\% & 9.39\% & 18.08\% \\
Christoffersen independence & 0.70\% & 11.03\% & 6.10\% & 14.79\% \\
Conditional coverage        & 3.53\% & 24.88\% & 29.23\% & 34.39 \% \\ 
Basel committee traffic light   & \colorbox{green!50}{69.1\%}\colorbox{yellow!50}{29.7\%}\colorbox{red!50}{0.5\%} &  \colorbox{green!50}{62.4\%}\colorbox{yellow!50}{25.9\%}\colorbox{red!50}{10.8\%} & N.A. & N.A. \\ \bottomrule
\end{tabular}
\caption{Summary statistics on coverage and independence tests for 1-day VaRs computed by the neural-SDE (nSDE) markert model and the filtered historical simulation (FHS) approach. For the four statistical tests, i.e. Kupiec POF test (two-sided), Kupiec POF test (one-sided), Christoffersen's independence test and conditional coverage mixed test, the table lists the percentage of the 852 tested option portfolios that fail each test at the 0.05 significance level.}
\label{tab:backtesting_mpor1}
\end{table}

\begin{figure}
    \centering
    \includegraphics[scale=.66]{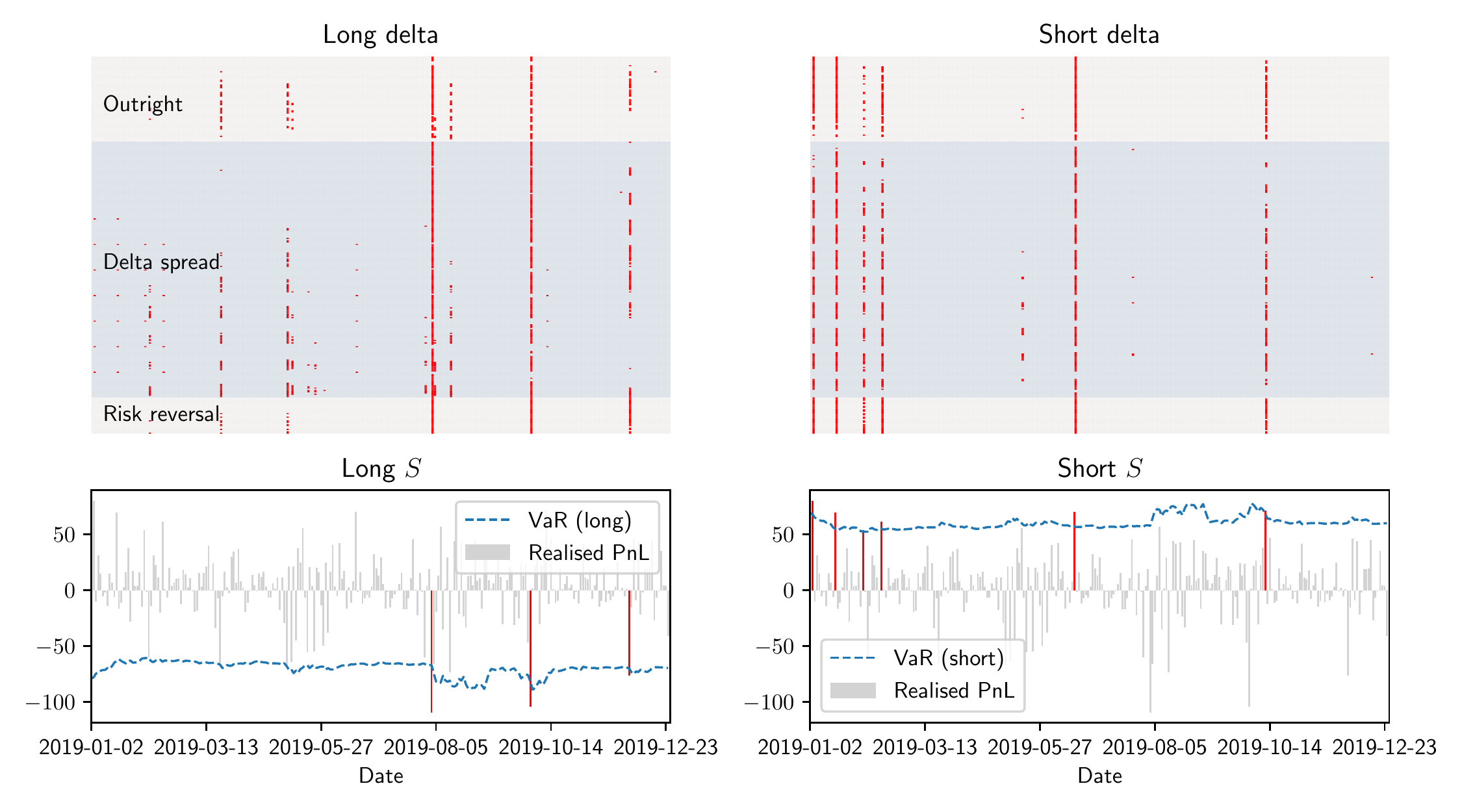}
    \caption{Times series of 1-day nSDE-$\text{VaR}_{0.99}$ breaches (red dots) for delta-exposed option portfolios, compared with that for positions in the underlying index. (EURO STOXX 50 index option)}
    \label{fig:VaR_breach_heatmap}
\end{figure}

To diagnose VaR coverage performances over different option portfolios, we group portfolios with different primary Greek-exposures, i.e. delta or vega, and examine whether their VaR breaches are synchronised during the testing window. In Figure \ref{fig:VaR_breach_heatmap}, we show the time series of 1-day nSDE-$\text{VaR}_{0.99}$ breaches for primarily delta-exposed option portfolios, and compare with positions in the underlying index. We see a few time points when VaR breaches happen simultaneously for multiple option portfolios, which correspond to synchronised VaR breaches or close-to-breaches in holding $S$.

\begin{figure}
    \centering
    \includegraphics[scale=.66]{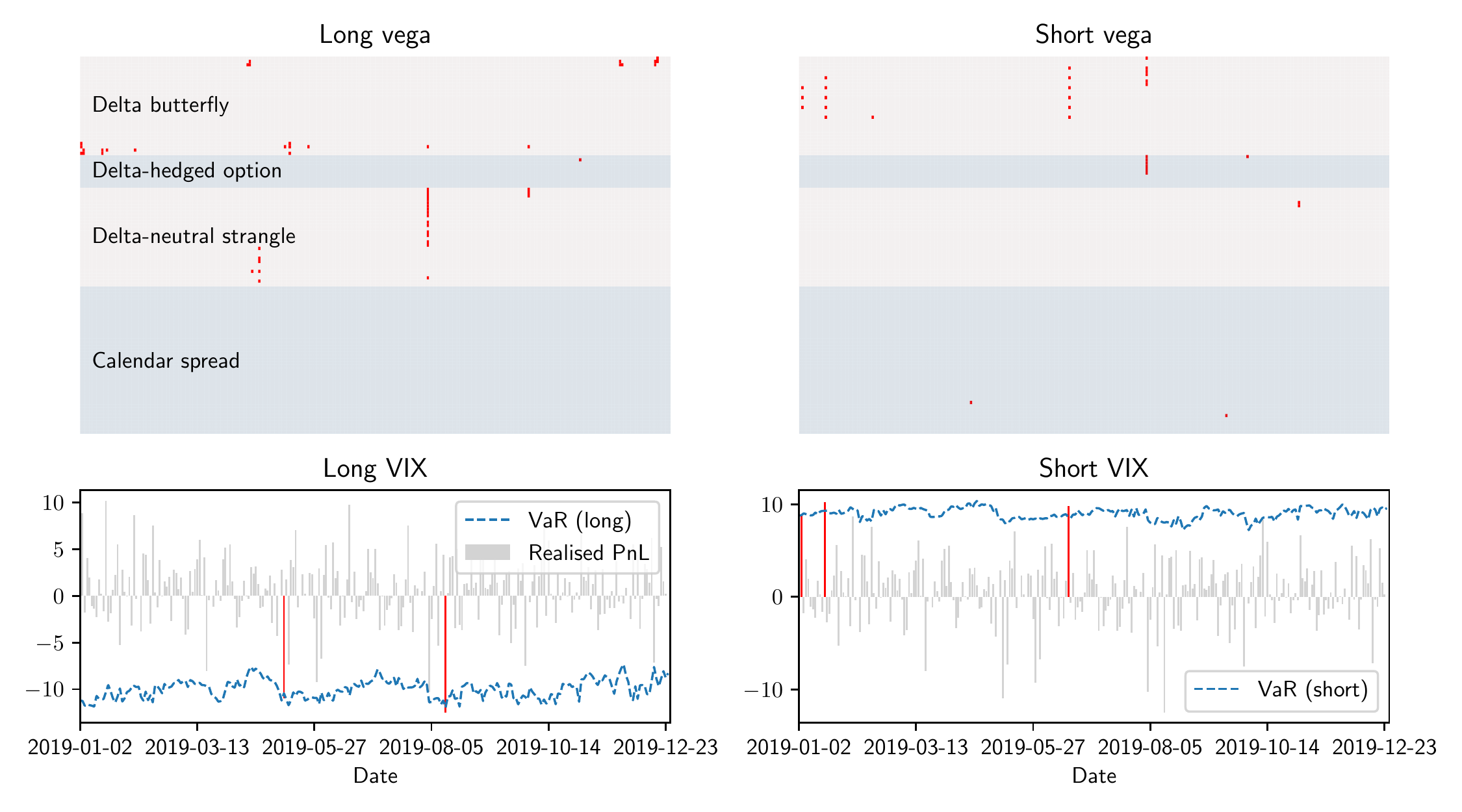}
    \caption{Times series of 1-day nSDE-$\text{VaR}_{0.99}$ breaches (red dots) for delta-neutral and vega-exposed option portfolios, compared with that for positions in the VIX portfolio.}
    \label{fig:VaR_breach_heatmap_dn}
\end{figure}

For delta-neutral and vega-exposed portfolios, we compare their VaR breaches with those of the VIX portfolio in Figure \ref{fig:VaR_breach_heatmap_dn}. There are few breaches observed, suggesting the VaR model is over conservative for these option types. We see some synchronised VaR breaches for vega-short positions in delta butterflies and the VIX. We perform a similar analysis for FHS-VaRs; as shown in Figure \ref{fig:VaR_breach_heatmap_dn_fhs} in Appendix \ref{apd:var_backtesting}, FHS-VaRs underestimate risks severely for most delta-hedged portfolios.

\begin{figure}[!ht]
    \centering
    \begin{subfigure}[b]{.49\textwidth}
    \centering
        \includegraphics[scale=.64]{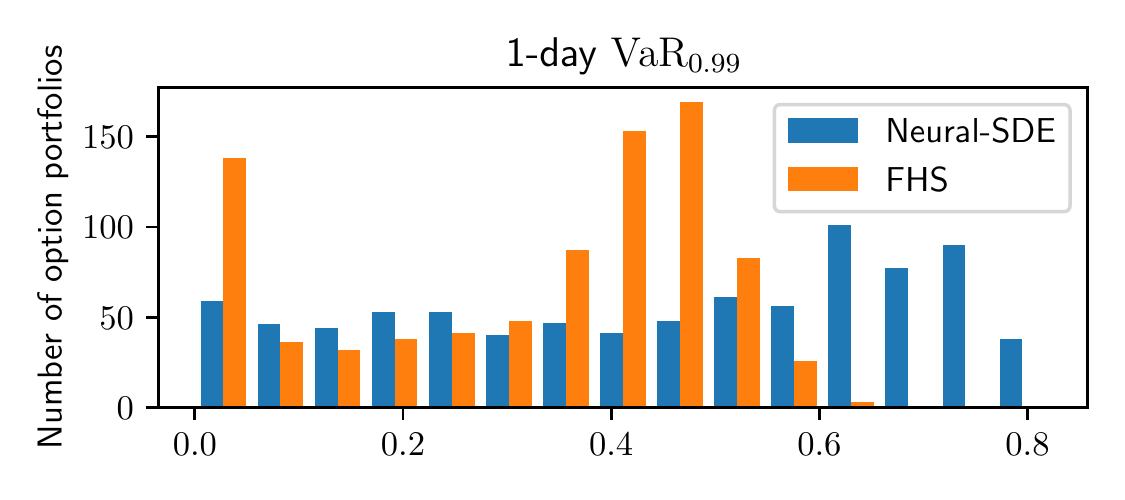}
    \end{subfigure}
    \hfill
    \begin{subfigure}[b]{.49\textwidth}
    \centering
        \includegraphics[scale=.64]{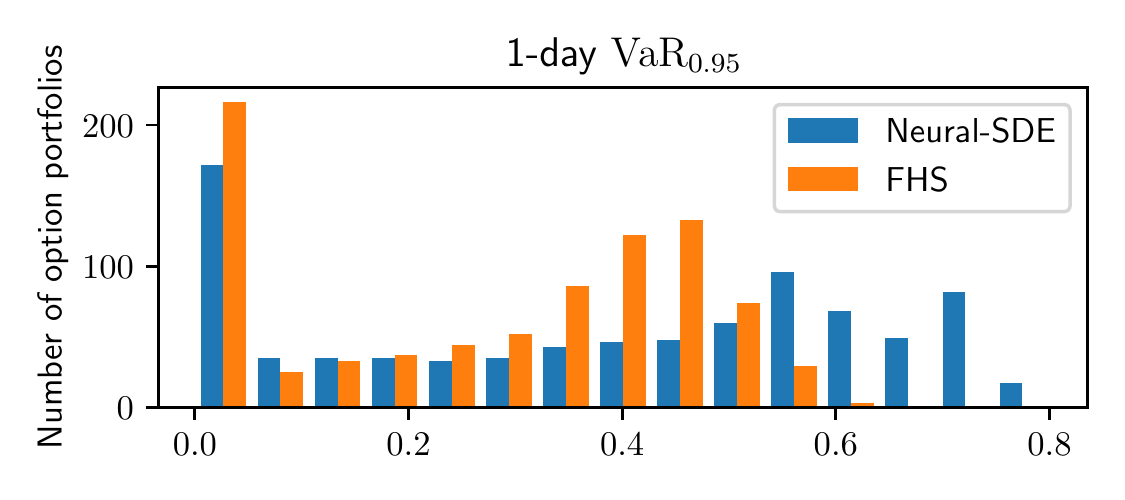}
    \end{subfigure}
    \caption{Distribution of trough-to-peak ratios (1-day VaRs) over different tested option portfolios.}
    \label{fig:ttp_mpor1}
\end{figure}

While coverage and independence tests assess whether a VaR calculation approach achieves desirable statistical properties, regulators would also like to avoid over-\textit{procyclical} VaR models. Procyclicality measures quantify variations in VaR values over the testing periods. Common risk models are procyclical because they estimate higher risk for the same portfolio in times of market stress and lower risk in calm markets. While procyclicality might indicate that a model is responsive to contemporary market volatility, it may also be an artifact of a poorly calibrated model. Excessive procyclicality can cause liquidity stress, where parties posting margin have to find additional liquid assets, often at times when it is most difficult for them to do so. We use the trough-to-peak ratio\footnote{The orginal paper \cite{pcc2014} uses peak-to-trough ratio,  the ratio of the maximum risk of a constant portfolio to the minimum risk over a fixed observation period. Here we take its reciprocal because for some heavily-hedged portfolio the minimum risk can be very close to 0, making the peak-to-trough ratio explode.} proposed by the Bank of England \cite{pcc2014} to measure procyclicality. After calculating the trough-to-peak ratio for each tested portfolio, we get a distribution of the ratio over our evaluated portfolios. In Figure \ref{fig:ttp_mpor1}, we compare the trough-to-peak distributions computed from the neural-SDE and the FHS approach. The neural-SDE approach tends to produce less procyclical 1-day VaRs at both 0.99 and 0.95 confidence levels. In particular, for 1-day $\text{VaR}_{0.99}$, there is a substantial reduction in the number of portfolios with very low trough-to-peak ratio (corresponding to a high level of procyclicality) and an increase in the number of ratios above 0.6. Consistent results for other risk horizons and confidence levels can be seen in Appendix \ref{apd:var_backtesting}.

While more backtesting results for 2-day, 5-day and 10-day VaRs and those for DAX options are reported in Appendix \ref{apd:var_backtesting}, here we briefly summarise the backtesting results for 10-day VaRs of EURO STOXX 50 index options in Table \ref{tab:backtesting_mpor10}\footnote{Given we consider breaches in overlapping 10-day periods, the Christoffersen independence test is not applicable.}. Similar to the observations of 1-day VaRs, the neural-SDE approach leads to less portfolios that fail the one-sided coverage test, the independence test and the conditional coverage test (the nulls are rejected at the 0.05 significance level for all tests). This indicates that the neural-SDE approach has consistently superior performance for long-horizon 0.99-confidence VaR calculation. For 0.95-confidence VaR, the neural-SDE approach is more conservative than the FHS approach, and the results are more mixed than in the 1-day case.

\begin{table}[!ht]
\footnotesize
\centering
\begin{tabular}{lcccc}
\toprule
\multirow{2}{*}{} & \multicolumn{2}{c}{10-day $\text{VaR}_{0.99}$} & \multicolumn{2}{c}{10-day $\text{VaR}_{0.95}$}\\ \cmidrule(lr){2-3} \cmidrule(lr){4-5}
                            & nSDE &  FHS & nSDE  & FHS \\ \cmidrule(lr){1-5}
Coverage ratio median       & 1.0000 & 0.9918 & 0.9712 & 0.9637 \\
Coverage ratio mean         & 0.9963 & 0.9800 & 0.9584 & 0.9464  \\
Kupiec PF (two-sided)      & 4.10\% & 17.92\% & 51.52\% & 37.12\%\\
Kupiec PF (one-sided)      & 4.10\% & 17.92\% & 10.07\% & 14.05\%  \\ \bottomrule
\end{tabular}
\caption{Summary statistics on coverage tests for 10-day VaRs computed by the neural-SDE (nSDE) market model and the filtered historical simulation (FHS) approach.}
\label{tab:backtesting_mpor10}
\end{table}

\subsection{Computation time comparison}

We have carried out the backtesting analysis on a 2.4GHz 8-core Intel Core i9-9980HK CPU with 32GB RAM, and summarised the average elapsed time for simulating one risk scenario per backtested day in Table \ref{tab:elapsed_time}.

\begin{table}[!ht]
\centering
\footnotesize
\begin{tabular}{lcccc}
\toprule
 & \multicolumn{2}{c}{Simulate \& revaluation} & \multicolumn{2}{c}{Repair arbitrage} \\ \cmidrule(lr){2-3} \cmidrule(lr){4-5}
 & nSDE & FHS (Heston) & nSDE & FHS (Heston) \\ \cmidrule(lr){1-5}
Expected time (ms) & 5.0 & 147.3 & 14.3 & N.A. \\
Std. time (ms) & 0.3 & 58.7 & 7.1 & N.A. \\
\bottomrule
\end{tabular}
\caption{Average and standard deviation of the elapsed time for simulation one risk scenario.}
\label{tab:elapsed_time}
\end{table}

In the FHS approach, evaluating option portfolios with simulated risk factor, i.e. the Heston parameters, is most time-consuming. We perform this by calling the efficiently implemented Matlab function \texttt{optByHestonNI} \cite{optByHestonNI}, which gives the option price by Heston model using numerical integration. In contrast, the nSDE approach evaluates option portfolios through a simple linear combination of simulated risk factors. This is the reason why the time for simulation and revaluation is reduced from 147.3 ms to 5.0 ms. The nSDE approach has an additional step of detecting and repairing arbitrage in the simulated option prices, which however is computationally cheap due to the efficient linear programming problem formulation in \cite{Cohen2020}.

\section{Conclusions and further extensions}
\label{sec:conclusion}

We estimate a neural-SDE market model using historical price data of EURO STOXX 50 and DAX index options, the top two most traded equity options listed on Eurex. We assess the in-sample goodness-of-fit quality by investigating the neural network training loss convergence, the post-fitting historical residuals, and the comparison between the volatility estimates for $\ln S$ with the benchmark GARCH(1,1) estimates. Thereafter, we verify that the trained model can forward simulate long trajectories of option price (or implied volatility) data that have similar marginal and joint distributions to historical data. By backtesting the option portfolio VaR generated from the trained neural-SDE market model in an out-of-sample setting, we find that the model beats the traditional filtered historical approaches with statistically more efficient coverage, less procyclicality and less computational efforts; and this is consistently true for VaRs computed with various risk horizons and confidence levels.

While seeing the success in applying the neural-SDE market model in simulation and risk management, there remain directions to extend the current framework for a wider range of applications. For example, a simple and straightforward extension is to examine other risk measures such as expected shortfall. Second, in order to model option types other than European vanillas, it is necessary to derive arbitrage constraints among those options, or to modify the approach taken accordingly.

\appendix
\small

\section{Data pre-processing}
\label{apd:data_cleansing}

Typical data vendors\footnote{For example, OptionMetrics (\url{https://wrds-www.wharton.upenn.edu/pages/about/data-vendors/optionmetrics/}), IVolatility (\url{https://www.ivolatility.com/home.j}), Nasdaq Data Link (\url{https://data.nasdaq.com/data/OWF-optionworks-futures-options/documentation}), Bloomberg, etc.} provide \emph{historical} EOD (end-of-day) option prices with parameters in the format of a fixed set of representative time-to-expiries and Black--Scholes deltas, instead of traded expiries and strikes. This format is popular in industry, as it saves users great efforts in building historical implied volatility surfaces, which otherwise requires the users themselves to reconcile traded option contracts that are changing over time.

This data format is also appealing to our modelling approach, though the raw data need to be pre-processed to (1) get rid of static arbitrage, and (2) interpolate prices to a fixed set of moneyness parameters. We now explain how we pre-process the OptionMetrics dataset as described in Section \ref{sec:option_data}.

\begin{figure}[!ht]
    \centering
    \includegraphics[scale=.66]{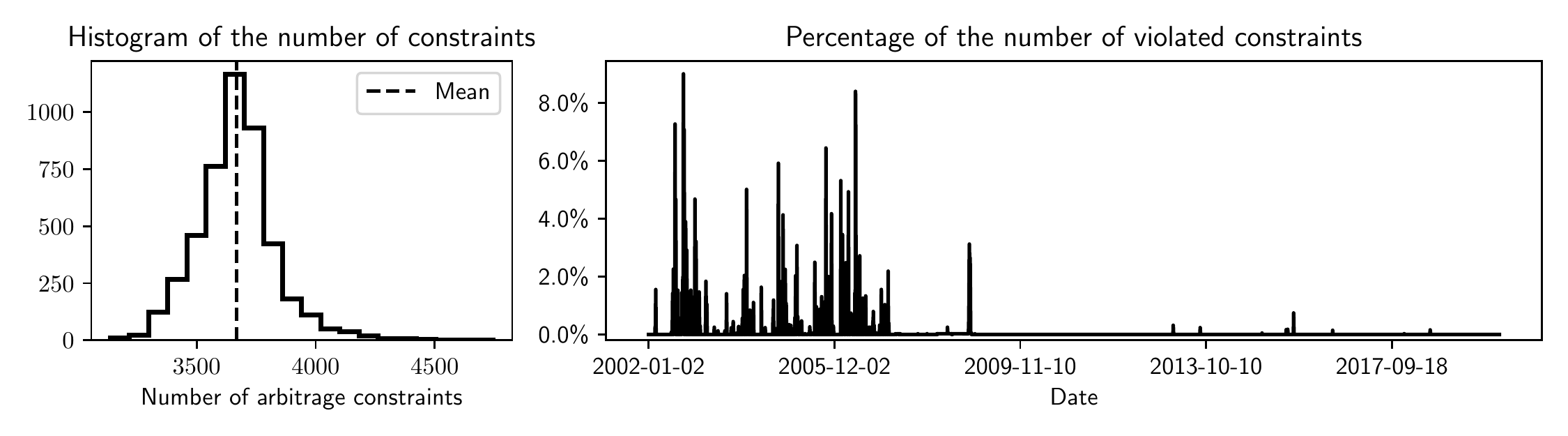}
    \caption{Statistics on the static arbitrage constraints violation for EURO STOXX 50 index option price data.}
    \label{fig:stats_arb_EX50}
\end{figure}

We use the arbitrage detection method of our previous work \cite{Cohen2020} to construct static arbitrage constraints in terms of option prices, and check the violation of these constraints by the data over time. In Figure \ref{fig:stats_arb_EX50}, we show the histogram for the number of constructed arbitrage constraints and what percentage of them are violated over time. On average, there are about 3668 constraints per day. Historical option prices are almost free of static arbitrage after October 2008, before which there were about $8 \sim 9\%$ of constraint violations, in the worst cases, in 2002 and 2006.

\begin{figure}[!ht]
    \centering
    \includegraphics[scale=0.66]{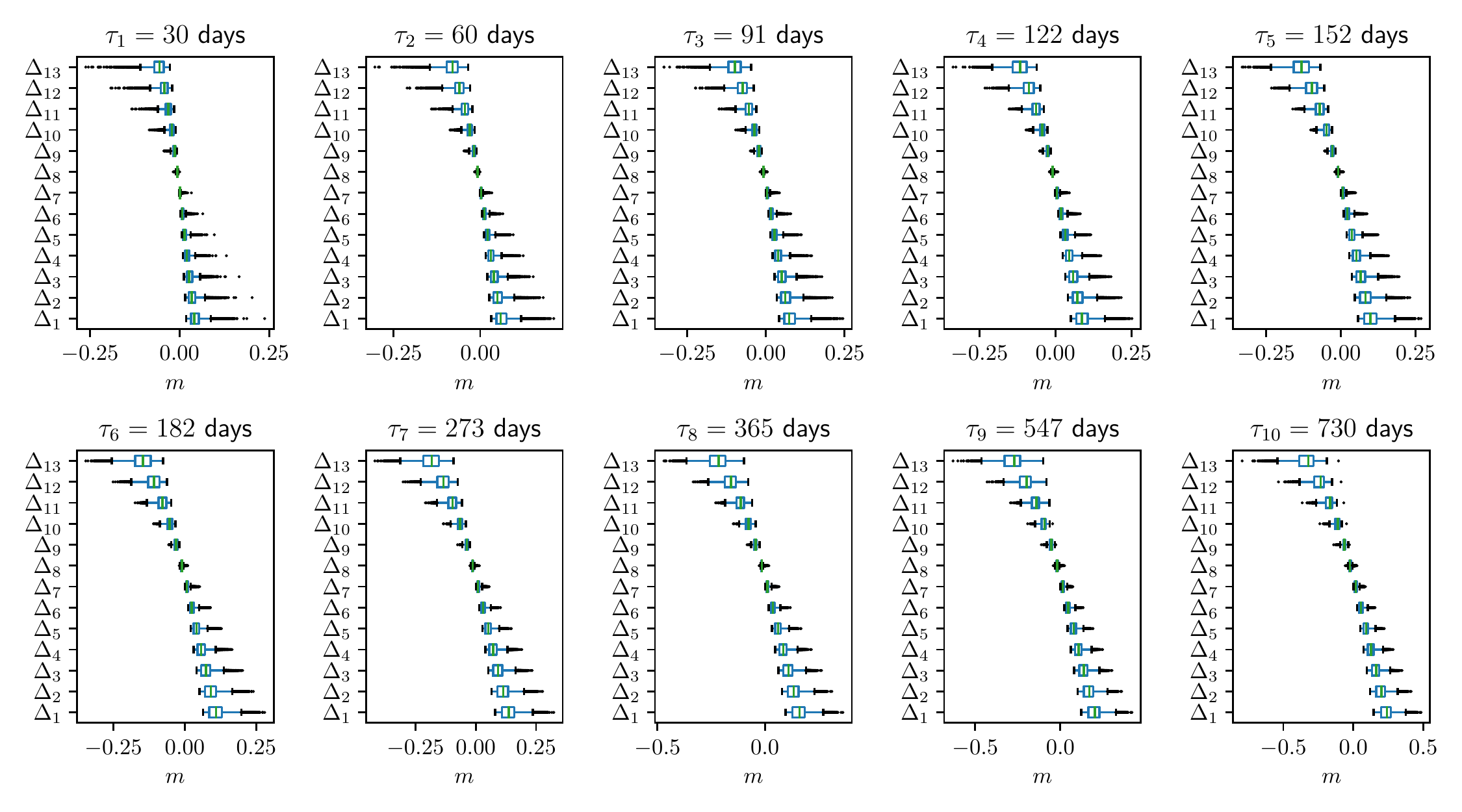}
    \caption{Boxplots of the moneynesses $m$ corresponding to the 13 deltas for the 10 time-to-expiries. $\Delta_1 = 0.2$, $\Delta_{13}=0.8$, and consecutive deltas have a $0.05$ increment.}
    \label{fig:median_ms}
\end{figure}

We use historical \emph{median} moneynesses to define our liquid option lattice $\mathcal{L}_\text{liq}$ (as shown in Figure \ref{fig:eurex_lattice}), and interpolate prices to the fixed set of moneyness parameters. In Figure \ref{fig:median_ms}, we show boxplots of moneynesses that are calculated over time for the fixed set of 13 deltas and 10 time-to-expiries. The closer to the money a delta is, the less variation we observe in the corresponding moneyness. This implies that less interpolation error is introduced to prices for close-to-the-money options, assuming a $C^2$ convex price curve over moneyness. We use cubic splines to interpolate the call option price curve over moneyness for each expiry.

\begin{figure}[!ht]
    \centering
    \includegraphics[scale=.66]{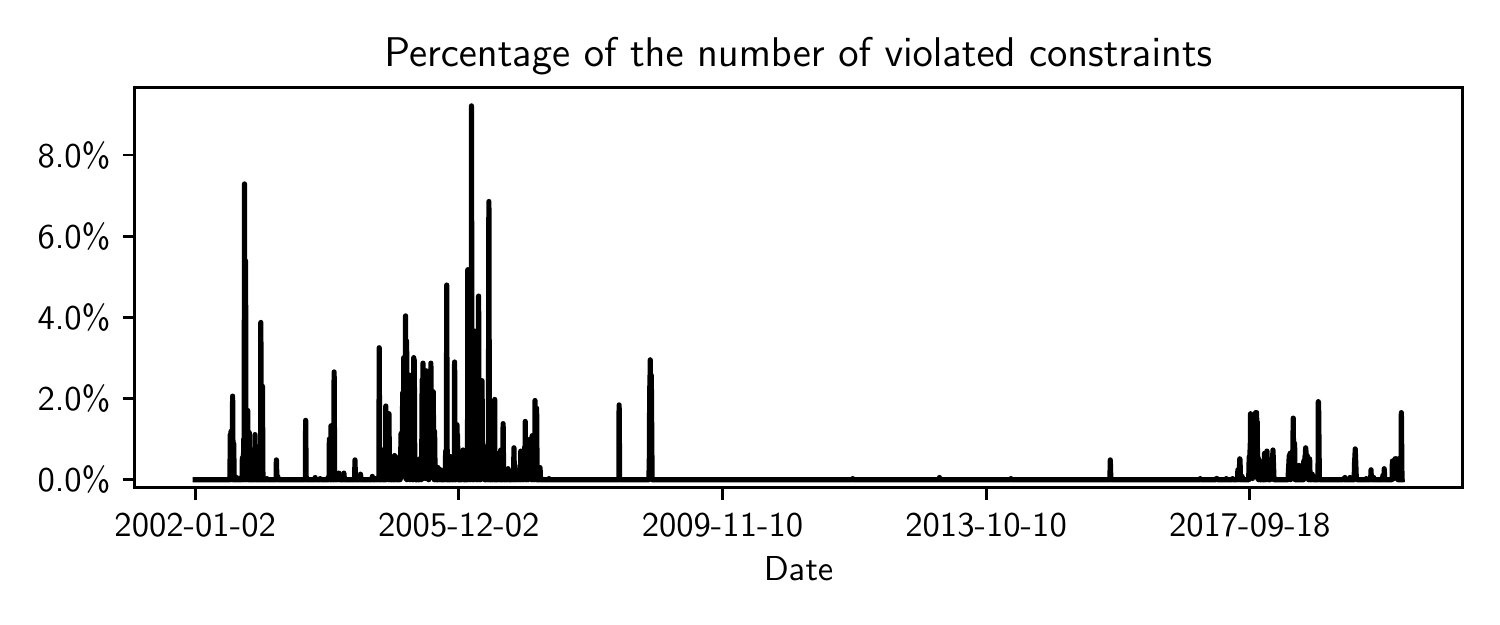}
    \caption{Statistics on the static arbitrage constraints violation for EURO STOXX 50 index option prices interpolated to the liquid lattice $\mathcal{L}_\text{liq}$.}
    \label{fig:stats_arb_interp_EX50}
\end{figure}

After interpolating option prices to the liquid lattice $\mathcal{L}_\text{liq}$ for each historical date, we check the amount of static arbitrage in the interpolated prices and present the fraction of arbitrage constraint violation in Figure \ref{fig:stats_arb_interp_EX50}. Compared with Figure \ref{fig:stats_arb_EX50}, the interpolation has introduced extra arbitrage for a few days, but in the worst case the introduced violation percentage is not greater than $2\%$. Thereafter, we perturb arbitrageable prices using the $\ell^1$-repair method \cite{Cohen2020} to get arbitrage-free prices.

\section{Alternative model configurations}

Apart from the market model discussed in the main text, we present two alternative models that differ in the number of primary factors chosen and the dependence between $S$ and $\xi$. We also show their corresponding VaR backtesting results in Appendix \ref{apd:var_backtesting}, which perform slightly worse than the model presented in the main text.

\subsection{A three-primary-factor market model}
\label{apd:3factor}

\begin{figure}[!ht]
    \centering
    \includegraphics[scale=.66]{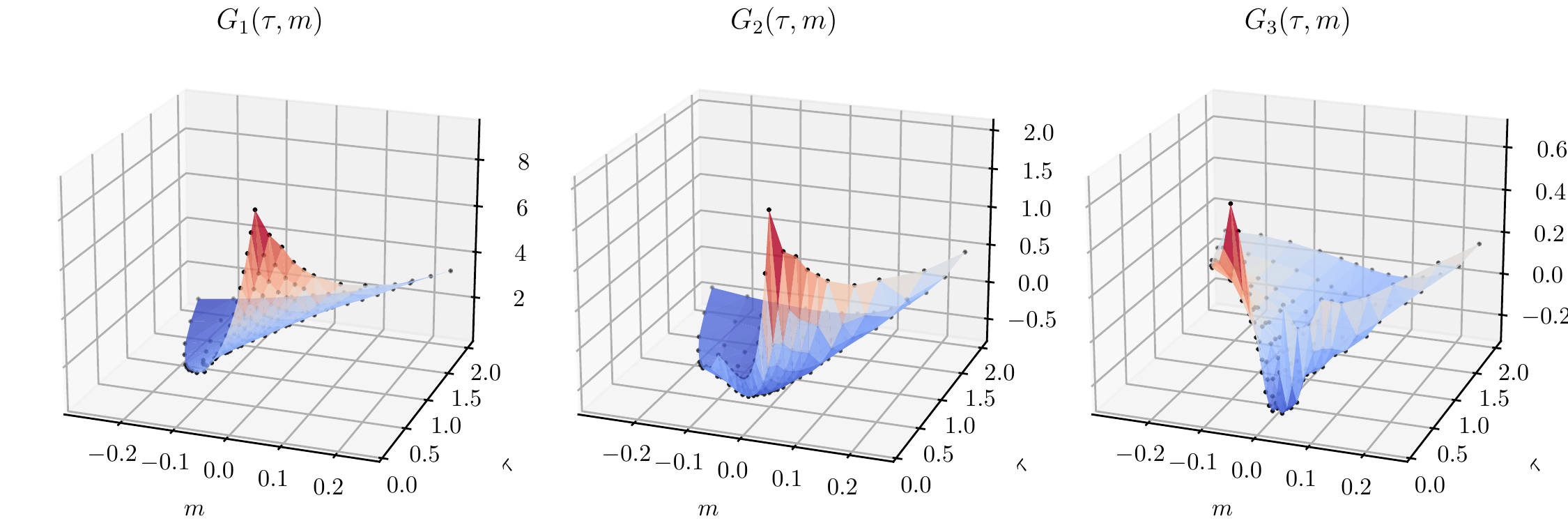}
    \caption{Price basis functions of the normalised call price surface.}
    \label{fig:Gs_primary_3factor}
\end{figure}

In Table \ref{tab:factor_metrics_3factor}, we display the types of the three primary factors $\xi = [\xi_1 ~\xi_2 ~\xi_3]^\top \in \mathbb{R}^3$ we use for representing call option prices, as well as the resultant MAPE, PDA and PSAS. Compared with the two primary factors (last row in Table \ref{tab:factor_metrics}), the additional dynamic arbitrage factor reduces dynamic arbitrage by 6.1\% and MAPE in price reconstruction error by 0.56\%, but introduces 0.55\% more static arbitrage. In Figure \ref{fig:Gs_primary_3factor}, we plot the corresponding price basis functions of these three factors.

\begin{table}[!ht]
    \centering
    \footnotesize
    \begin{tabular}{lccc}
        \toprule
        Factors & MAPE & PDA & PSAS  \\
        \cmidrule(lr){1-1} \cmidrule(lr){2-4}
        Dynamic arb. + Statistical acc. + Static arb. & $4.26\%$ & $1.73\%$ & $0.85\%$ \\
        \bottomrule
    \end{tabular}
    \caption{MAPE, PDA and PSAS metrics when including three primary factors.}
    \label{tab:factor_metrics_3factor}
\end{table}

\begin{figure}[!ht]
    \centering
    \includegraphics[scale=0.66]{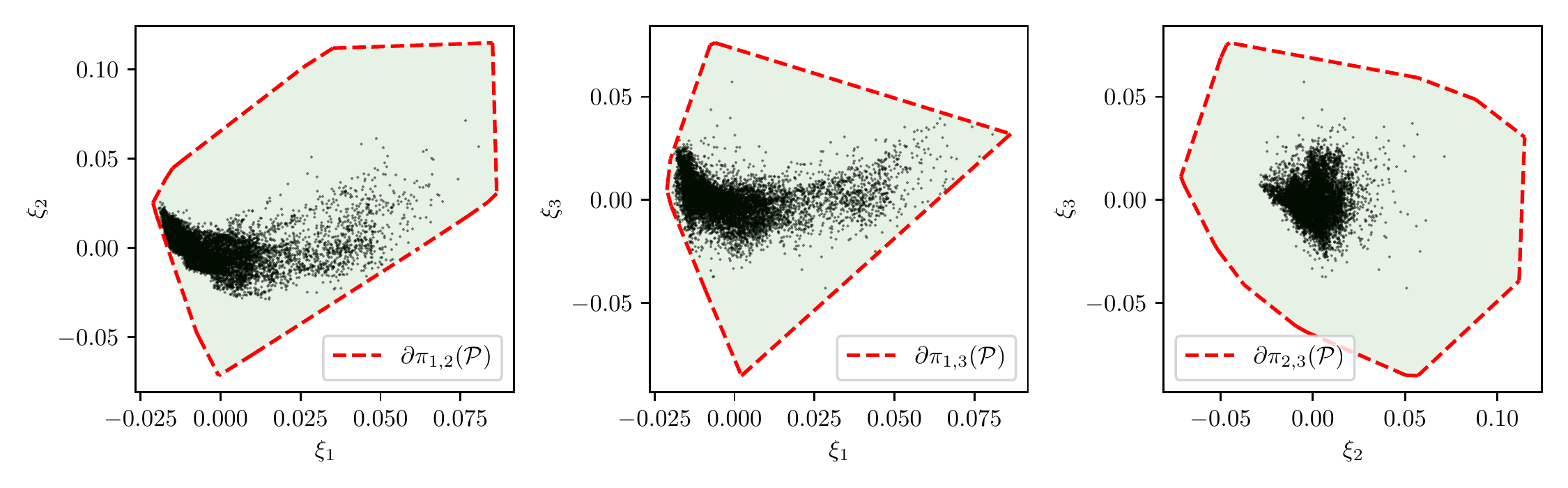}
    \caption{Trajectory (black dots) of the primary $\mathbb{R}^3$ factors and the corresponding static arbitrage constraints (red and orange dashed lines), projected onto $\mathbb{R}^2$ spaces.}
    \label{fig:factor3}
\end{figure}

In Figure \ref{fig:factor3}, the black dots give trajectories of $(\xi_i, \xi_j)$ for $i,j \in \{1,2,3\}$. For each $(i,j)$, we also indicate boundaries of the polygon, that is, the 2D affine projection of the 3D arbitrage-free state space for $\xi$. Specifically, let $\mathcal{P}$ be the $\mathbb{R}^3$-polytope state space for the factors $\xi$ formed by the static arbitrage constraints, then the affine projection of this polytope onto the $(\xi_i, \xi_j)$-space is given by
\begin{equation*}
    \pi_{i,j}(\mathcal{P}) = \left\{ y \in \mathbb{R}^2 : \exists \xi = (\xi_k) \in \mathcal{P}, y = [\xi_i  ~\xi_j]^\top \right\}.
\end{equation*}
We plot the boundaries $\partial \pi_{i,j}(\mathcal{P})$ as red dashed lines in Figure \ref{fig:factor3}. These boundaries (which are, in some sense, as wide as possible) suggest how tight the arbitrage bounds are.

\begin{figure}[!ht]
    \centering
    \includegraphics[scale=0.66]{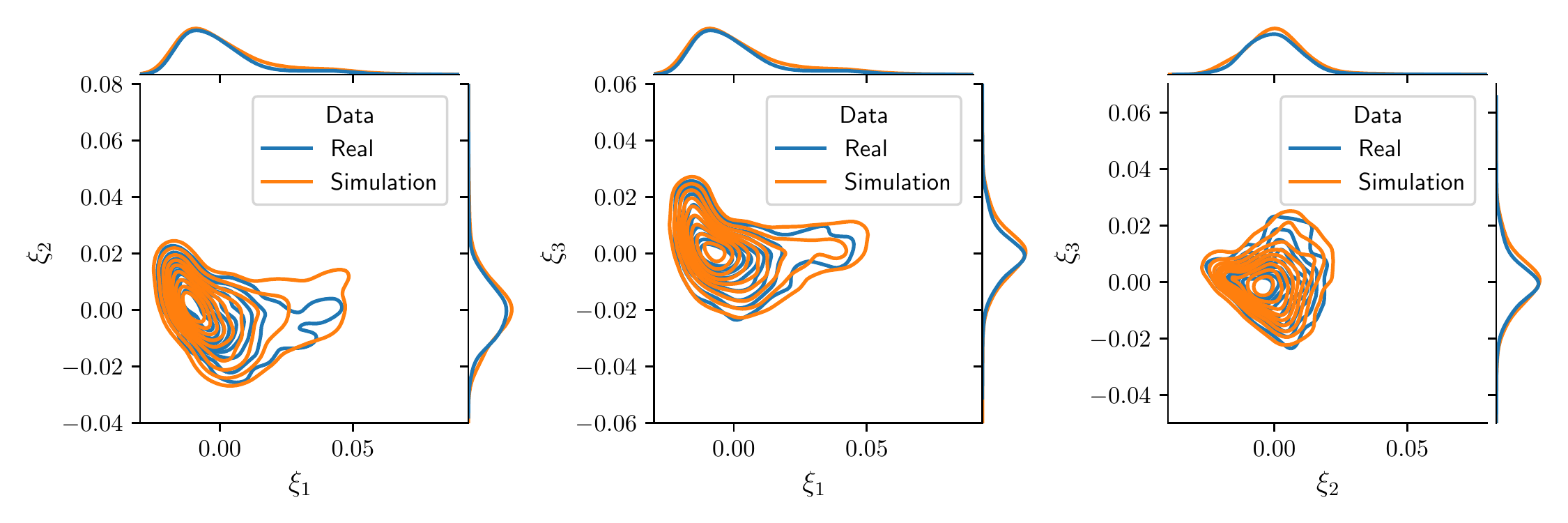}
    \caption{Joint distribution of pairs of the simulated primary factors, compared with that of the real data.}
    \label{fig:factor3_sim}
\end{figure}

We take the three factors and model their dynamics using neural-SDEs. After training the models, we forward-simulate $10,000$ timesteps, and compare the pairwise joint distribution of the simulated factors and that of the real data, as shown in Figure \ref{fig:factor3_sim}. The trained neural-SDE model seems to be capable of simulating long trajectory of the three factors that are similar to the real data, at least in terms of pairwise joint distribution.

\subsection{Modelling and estimating $S$ and $\xi$ jointly}
\label{apd:model_xiS}

We build and estimate an SDE model for $\tilde{\xi} = (\ln S, \xi)$ with a fully-specified diffusion matrix; in fact, the model \eqref{eq:market_model} is a special case with a block-diagonal diffusion matrix. Given $\xi \in \mathbb{R}^d$, compared with the model \eqref{eq:market_model}, the joint model has $d$ more $\mathbb{R}$-valued functions to estimate in the diffusion term, which captures dynamic correlation between $\ln S$ and each of the $d$ factors.

\begin{figure}[!ht]
    \centering
    \includegraphics[scale=.66]{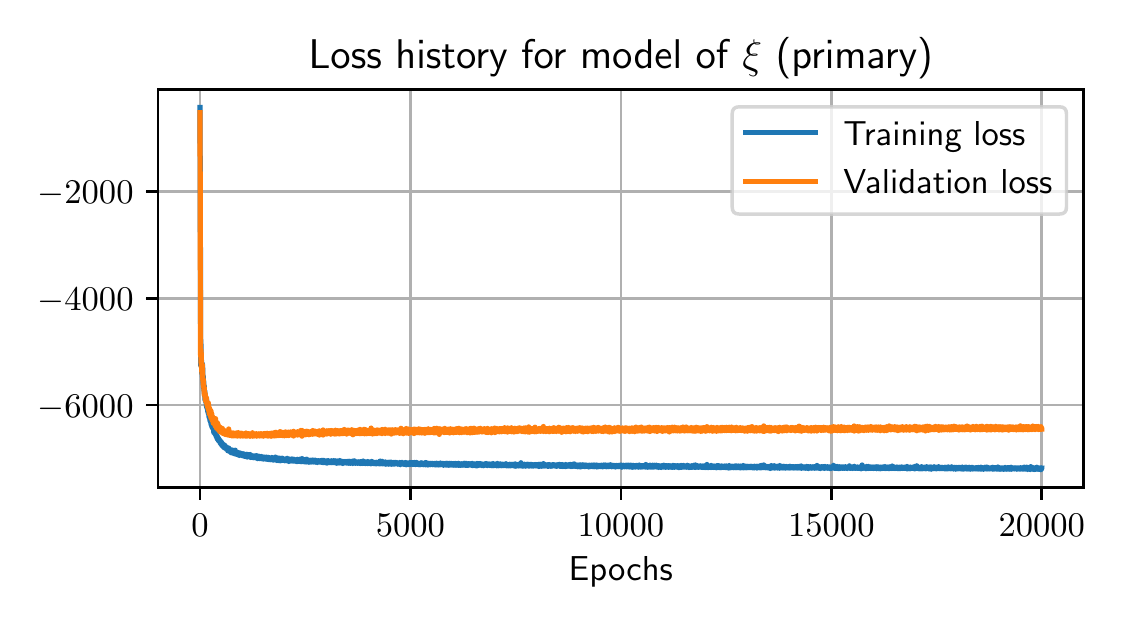}
    \caption{Evolution of training losses and validation losses for the joint model of $\ln S$ and $\xi$.}
    \label{fig:loss_hist_xiS}
\end{figure}

For estimating the joint model, we still use the concatenated data (of EURO STOXX 50 and DAX options) for training the neural nets. We assume that the two stock index processes have the same diffusion function with respect to $\xi$ but different constant drifts estimated by \eqref{eq:S_drift_est}. This enables us to use the concatenated stock index data to estimate a universal volatility function for the two indices. In Figure \ref{fig:loss_hist_xiS}, we show the loss history of training the neural nets. Both the training and validation losses decline rapidly over the first 1000 epochs and then converge gradually.

\begin{figure}[!ht]
    \centering
    \includegraphics[scale=.66]{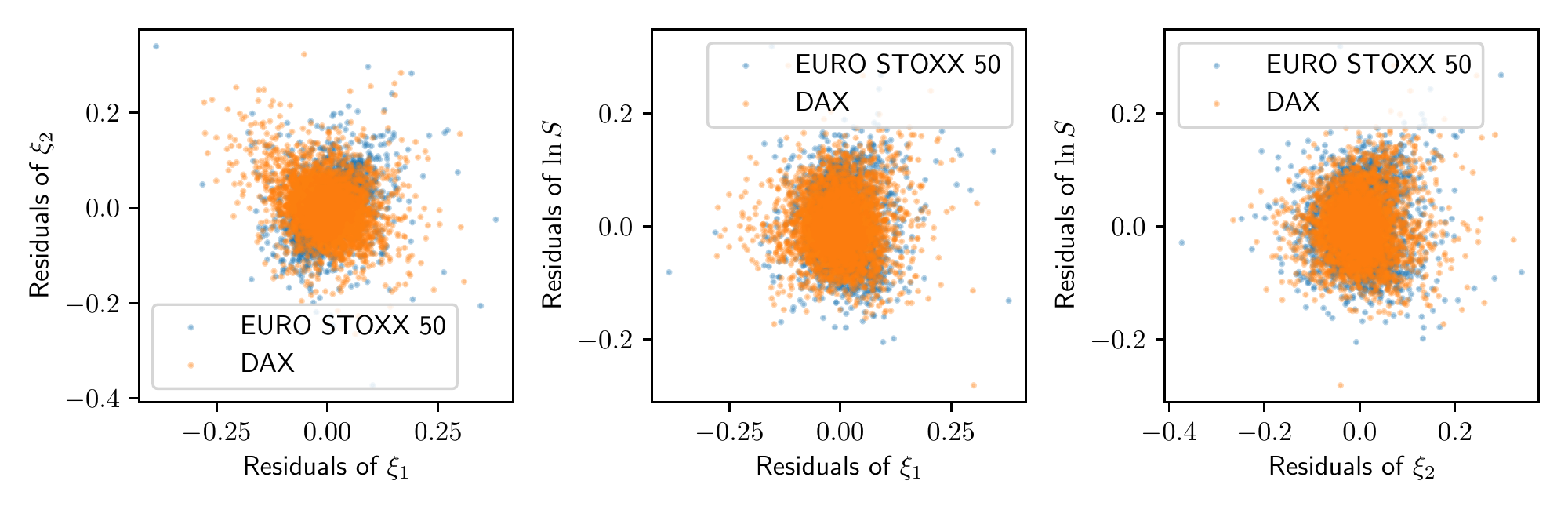}
    \caption{Scattter plots of historical in-sample model residuals.}
    \label{fig:hist_res_jointdensity_xiS}
\end{figure}

To see that the joint model has reasonably captured covariances between the underlying and the the factors, we show in Figure \ref{fig:hist_res_jointdensity_xiS} the scatter plots of all pairs of model residuals. Compared with Figure \ref{fig:hist_res_jointdensity}, the joint model takes into account the leverage effect and leaves no correlation between residuals of $\ln S$ and $\xi_1$.

\section{Heston model calibration results and historical scenarios}
\label{apd:heston_calib}

\begin{figure}[!ht]
    \centering
    \includegraphics[scale=.66]{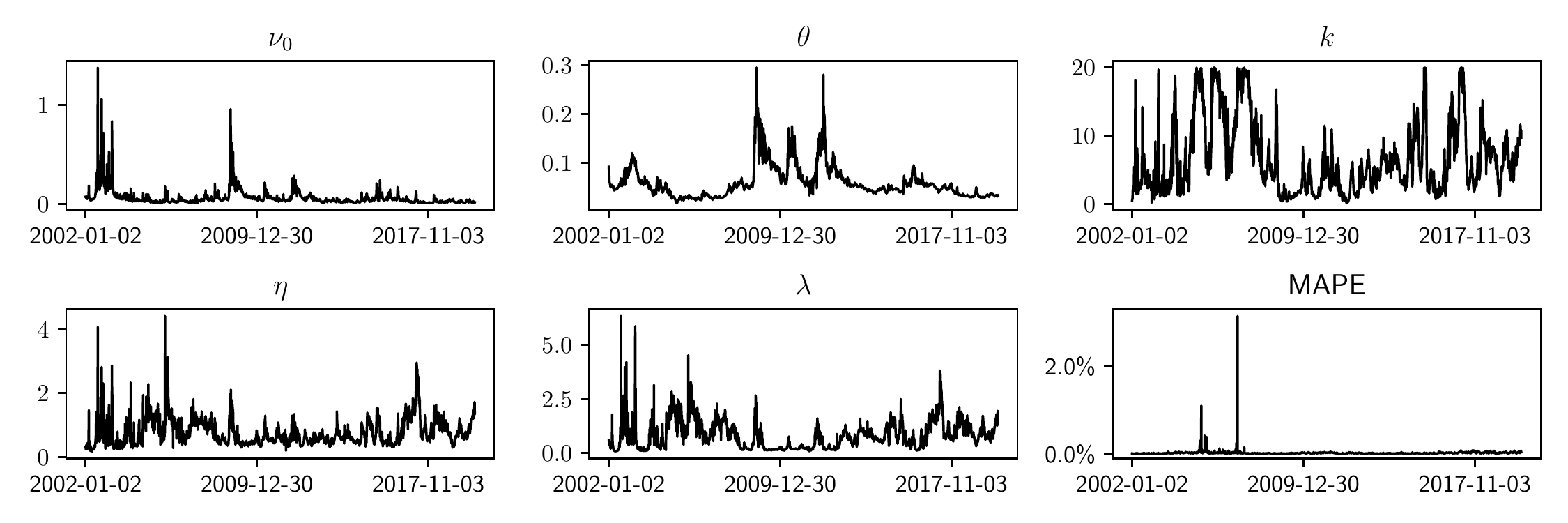}
    \caption{Heston model calibrated parameters and calibration errors (measured as mean absolute percentage error) for historical price data of EURO STOXX 50 index options.}
    \label{fig:heston_calib_EX50}
\end{figure}

Heston models are constructed from the SDE
\begin{subequations}
\begin{align}
    \diff S_t & = r_t S_t \diff t + \sqrt{\nu_t} S_t \diff W_t^S, \\
    \diff \nu_t & = k (\theta - \nu_t) \diff t + \sigma \sqrt{\nu_t} \diff W_t^\nu, ~
    \diff \langle W_t^S, W_t^\nu \rangle = \rho \diff t, \label{eq:heston_nu}
\end{align}
\end{subequations}%
We re-write the variance process \eqref{eq:heston_nu} as
\begin{equation*}
    \diff \nu_t = k (\theta - \nu_t) \diff t + \eta \sqrt{\nu_t} \diff W_t^S + \lambda \sqrt{\nu_t} \diff W_t^\perp, \quad \diff \langle W_t^S, W_t^\perp \rangle = 0,
\end{equation*}
where $\eta = -\sigma \rho$, $\lambda = \sigma \sqrt{1-\rho^2}$. We characterise option price surfaces by the parameters $\Theta = (S_0, \nu_0, \theta, k, \eta, \lambda)$. On each historical date $t$, we calibrate parameters for a Heston model by solving
\begin{equation*}
    \widehat{\Theta}_t = \underset{\Theta}{\arg\min} \sum_{j=1}^N \frac{1}{\mathcal{V}_t(\tau_j, m_j)} \left( c^\Theta (\tau_j, m_j) - \tilde{c}_t (\tau_j, m_j) \right)^2,
\end{equation*}
where $\mathcal{V}$ is the option vega calculated through \eqref{eq:vega}. In Figure \ref{fig:heston_calib_EX50}, we show the time series of the calibrated parameters and calibration errors, measured by MAPE, for historical price data of EURO STOXX 50 index options. Since $\nu_0$, $\theta$, $k$ and $\lambda$ are positive by construction, we include the positivity constraint for these parameters during the calibration. While $\eta$ can be positive or non-positive, the calibrated result is always positive, implying negative correlations $\rho$ between the index and volatility over time.

\begin{figure}[!ht]
    \centering
    \includegraphics[scale=.66]{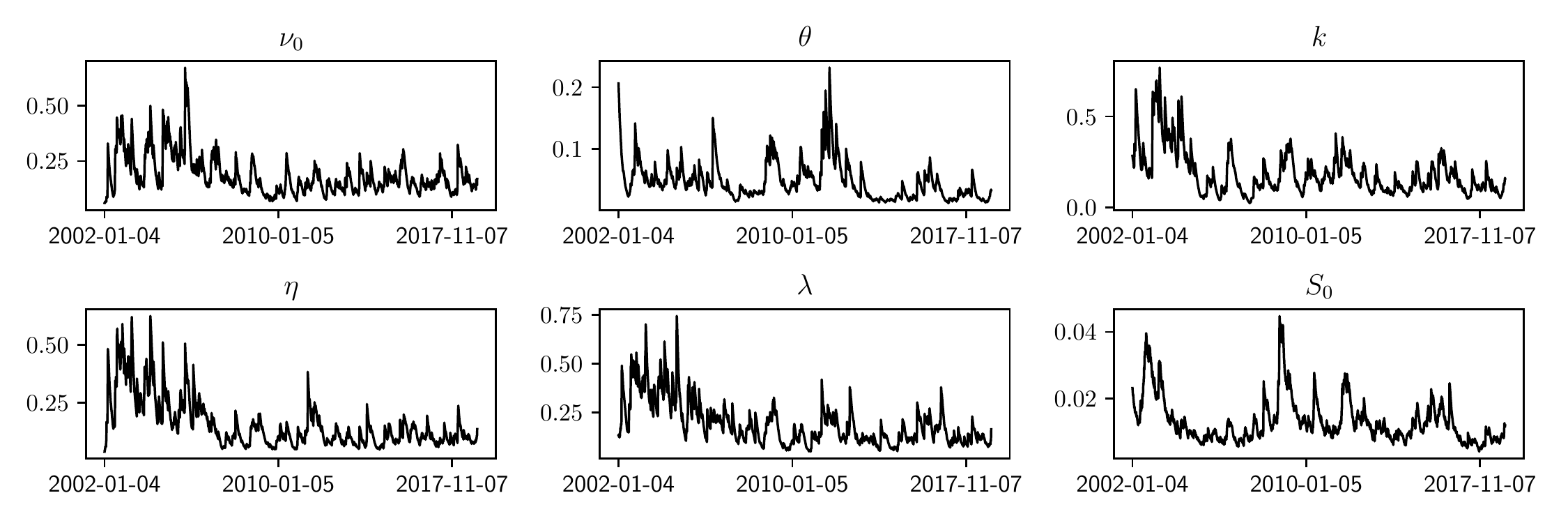}
    \caption{EWMA 1-day volatility estimates (with time decay factor $0.95$) for the Heston risk factors for EURO STOXX 50 index options.}
    \label{fig:heston_ewma_EX50}
\end{figure}

\begin{figure}[!ht]
    \centering
    \includegraphics[scale=.66]{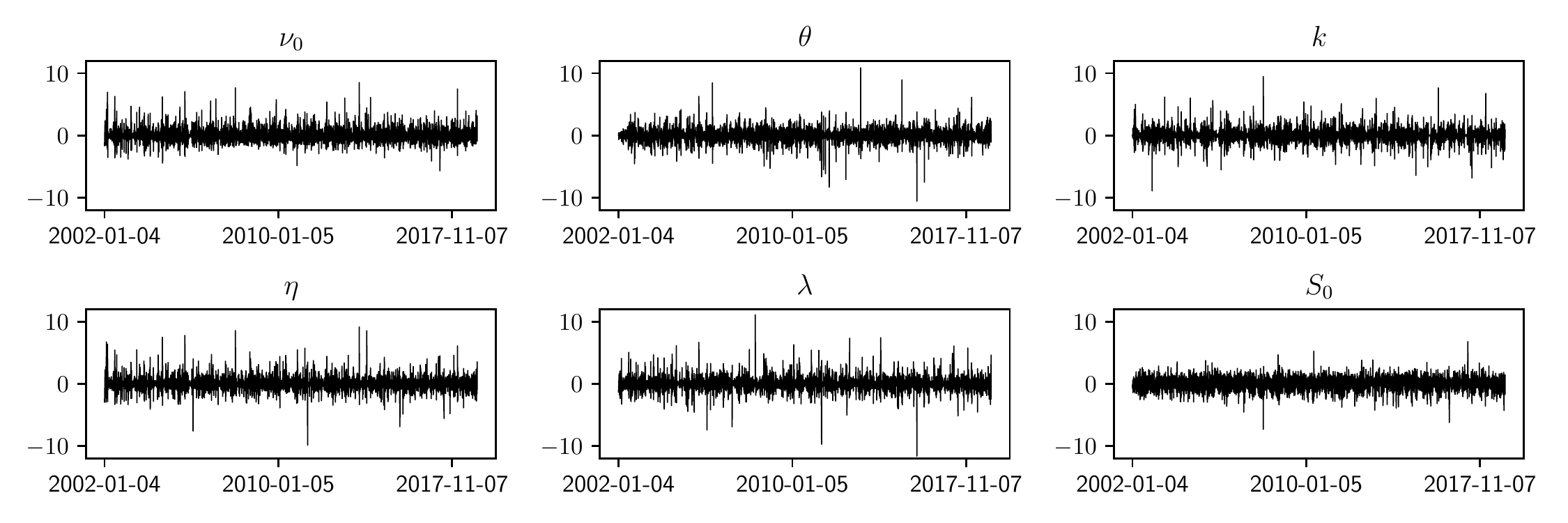}
    \caption{Standardised residuals of the Heston risk factors for EURO STOXX 50 index options.}
    \label{fig:heston_stdres_EX50}
\end{figure}

Given that all risk factors are positive, we compute their $1$-day log-returns, based on which we estimate the exponentially weighted moving average (EWMA) volatilities with time decay factor $0.95$. We plot the EWMA $1$-day volatility estimates over the estimation window in Figure \ref{fig:heston_ewma_EX50}. The RiskMetrics Technical document \cite{j1996riskmetrics} suggests a value of $0.94$ for the decay factor when estimating daily volatility. Nevertheless, in order to make the FHS-VaR and the nSDE-VaR more comparable, we optimise the choice of the decay factor by minimising the Wasserstein distance between the resulting EWMA vol and the in-sample neural-SDE vol in $S$ (as shown in Figure \ref{fig:vol_S}). This yields the decay factor 0.95. Dividing the returns by the volatility estimates, we obtain historical standardised residuals, or \emph{historical shocks}, as shown in Figure \ref{fig:heston_stdres_EX50}. The standardised residuals appear to be roughly stationary over time and do not exhibit volatility clustering. In order to generate risk scenarios for each $t$ in the testing window, we multiply these historical shocks by the time-$t$ $1$-day EWMA volatility, as described in \eqref{eq:filtered_return}, and apply these filtered returns to the time-$t$ risk factors, as if these historical shocks actually occurred.

\section{Tested option portfolios for risk simulation}
\label{apd:test_strategy}

To assess performance of a risk model, we backtest its risk evaluation of a variety of representative trading strategies. In particular, defining a set of deltas $\mathcal{D} = \{0.2, 0.3, 0.4, 0.5, 0.6, 0.7, 0.8\}$ and a set of time-to-expiries $\mathcal{T} = \{30, 60, 91, 122, 152, 182, 273, 365, 547, 730\} $ (in calendar days), we consider:

\begin{enumerate}[label=(\arabic*)]
    \item \emph{Outright call} option $C(\tau, \Delta)$ with all deltas $\Delta \in \mathcal{D}$ and all time-to-expiries $\tau \in \mathcal{T}$. There are in total $7 \times 10 = 70$ outright test strategies.
    \item \emph{Delta spread} $C(\tau, \Delta_1) - C(\tau, \Delta_2)$ with all distinguished delta pairs $\Delta_1 > \Delta_2 \in \mathcal{D}$ and all time-to-expiries $\tau \in \mathcal{T}$. There are in total ${7 \choose 2} \times 10 = 210$ delta spread test strategies.
    \item \emph{Delta butterfly} $C(\tau, \Delta) + C(\tau, 1-\Delta) - 2 \times C(\tau, 0.5)$ with all deltas $\Delta \in \mathcal{D} \setminus \{0.5\}$ and all time-to-expiries $\tau \in \mathcal{T}$. There are in total $3 \times 10 = 30$ delta butterfly spread test strategies.
    \item \emph{Delta-hedged option} $C(\tau, \Delta) - \Delta \times S$ with $\Delta = 0.5$ and all time-to-expiries $\tau \in \mathcal{T}$. There are in total $1 \times 10 = 10$ delta-hedged option test strategies.
    \item \emph{Delta-neural strangle} $C(\tau, \Delta) + C(\tau, 1 - \Delta) - S$ with all deltas $\Delta \in \mathcal{D} \setminus \{0.5\}$ and all time-to-expiries $\tau \in \mathcal{T}$. There are in total $3 \times 10 = 30$ delta-neutral strangle test strategies.
    \\Note that a typical delta-neutral straggle involves simultaneously buying a call and a put, i.e. $C(\tau, \Delta) + P(\tau, -\Delta)$. Here we use put-call parity to express such straddles in terms of only calls and the underlyings, ignoring cash components.
    \item \emph{Risk reversal} $C(\tau, \Delta) - C(\tau, 1 - \Delta) + S$ with all deltas $\Delta \in \{0.2, 0.3, 0.4\}$ and all time-to-expiries $\tau \in \mathcal{T}$. There are in total $3 \times 10 = 30$ risk reversal test strategies.
    \\Note that a typical risk reversal consists of buying an OTM call and selling an OTM put, i.e. $C(\tau, \Delta) - P(\tau, -\Delta)$. Here we use put-call parity to express such straddles in terms of only calls and the underlyings, ignoring cash components.
    \item \emph{Calendar spread} $C(\tau_1, \Delta) - C(\tau_2, \Delta)$  with $\Delta = 0.5$ and all distinguished time-to-expiry pairs $\tau_1 > \tau_2 \in \mathcal{T}$. There are in total $1 \times {10 \choose 2} = 45$ calendar spread test strategies.
    \item \emph{VIX}, which is a linear combination of OTM call and put option prices, and can be further written as a linear combination of call prices only, provided that put-call parity holds under no-arbitrage; see details in \cite{vix}.
\end{enumerate}

In total, as we consider both long and short positions, there are $ = 2 \times (70 + 210 + 30 + 10 + 30 + 30 + 45 + 1) = 852$ tested trading strategies. Among the tested trading strategies, delta butterflies, delta-hedged options, delta-neutral strangles and calendar spreads have low values of Black--Scholes delta.

\section{More VaR backtesting results}
\label{apd:var_backtesting}

We show backtesting results, using the model in the main text, for 2-day, 5-day and 10-day VaRs for EURO STOXX 50 index option portfolios in Table \ref{tab:backtesting_mpor2}. Similarly, we show backtesting results for 1-day, 2-day, 5-day and 10-day VaRs for DAX option portfolios in Table \ref{tab:backtesting_mpor_dax}. Aligned with our observation in the main text, the nSDE-VaR model outperforms the FHS-VaR approach through the backtesting analysis in almost all metrics.

\begin{table}[!ht]
\scriptsize
\centering
\begin{tabular}{lcccccccc}
\toprule
\multirow{2}{*}{}                             & \multicolumn{2}{c}{2-day $\text{VaR}_{0.99}$} & \multicolumn{2}{c}{2-day $\text{VaR}_{0.95}$} & \multicolumn{2}{c}{5-day $\text{VaR}_{0.99}$} & \multicolumn{2}{c}{5-day $\text{VaR}_{0.95}$} \\ \cmidrule(lr){2-3} \cmidrule(lr){4-5} \cmidrule(lr){6-7} \cmidrule(lr){8-9}
                            & nSDE &  FHS & nSDE  & FHS & nSDE & FHS & nSDE  & FHS \\ \cmidrule(lr){1-9}
Coverage ratio median       & 0.9960 & 0.9920 & 0.9602 & 0.9602 & 0.9960 & 0.9879 & 0.9677 & 0.9637 \\
Coverage ratio mean         & 0.9950 & 0.9800 & 0.9602 & 0.9402 & 0.9944 & 0.9798 & 0.9618 & 0.9464 \\
Kupiec PF (two-sided)      & 0.12\% & 13.11\% & 38.29\% & 35.71\% & 1.41\% & 12.76\% & 45.08\% & 37.12\% \\
Kupiec PF (one-sided)      & 0.12\% & 13.11\% & 7.26\% & 15.69\% & 1.41\% & 12.76\% & 7.61\% & 14.05\% \\ \bottomrule
\end{tabular}
\caption{Summary statistics on coverage tests for 2-day and 5-day VaRs computed by the neural-SDE (nSDE) market model and the filtered historical simulation (FHS) approach. These results are for EURO STOXX 50 index options.}
\label{tab:backtesting_mpor2}
\end{table}

\begin{table}[!ht]
\scriptsize
\centering
\begin{tabular}{lcccc}
\toprule
\multirow{2}{*}{}                             & \multicolumn{2}{c}{1-day $\text{VaR}_{0.99}$} & \multicolumn{2}{c}{1-day $\text{VaR}_{0.95}$} \\ \cmidrule(lr){2-3} \cmidrule(lr){4-5}
                            & nSDE &  FHS & nSDE  & FHS \\ \cmidrule(lr){1-5}
Coverage ratio median       & 0.9881 & 0.9841 & 0.9484 & 0.9524 \\
Coverage ratio mean         & 0.9872 & 0.9726 & 0.9556 & 0.9314 \\
Kupiec PF (two-sided)      & 14.64\% & 19.56\% & 33.49\% & 34.19\% \\
Kupiec PF (one-sided)      & 14.64\% & 19.56\% & 6.21\% & 18.03\% \\
Christoffersen independence & 0.47\% & 7.61\% & 4.68\% & 10.42\% \\
Conditional coverage        & 8.31\% & 19.44\% & 22.60\% & 33.49 \% \\ 
Basel committee traffic light   & \colorbox{green!50}{67.3\%}\colorbox{yellow!50}{31.1\%}\colorbox{red!50}{0.8\%} &  \colorbox{green!50}{56.9\%}\colorbox{yellow!50}{28.7\%}\colorbox{red!50}{12.9\%} & N.A. & N.A. \\ \bottomrule
\end{tabular}
\begin{tabular}{lcccccccc}
\toprule
\multirow{2}{*}{}                             & \multicolumn{2}{c}{2-day $\text{VaR}_{0.99}$} & \multicolumn{2}{c}{2-day $\text{VaR}_{0.95}$} & \multicolumn{2}{c}{5-day $\text{VaR}_{0.99}$} & \multicolumn{2}{c}{5-day $\text{VaR}_{0.95}$} \\ \cmidrule(lr){2-3} \cmidrule(lr){4-5} \cmidrule(lr){6-7} \cmidrule(lr){8-9}
                            & nSDE &  FHS & nSDE  & FHS & nSDE & FHS & nSDE  & FHS \\ \cmidrule(lr){1-9}
Coverage ratio median       & 0.9960 & 0.9920 & 0.9681 & 0.9482 & 0.9960 & 0.9839 & 0.9758 & 0.9637 \\
Coverage ratio mean         & 0.9903 & 0.9811 & 0.9602 & 0.9335 & 0.9928 & 0.9778 & 0.9644 & 0.9453 \\
Kupiec PF (two-sided)      & 8.31\% & 13.93\% & 54.33\% & 37.00\% & 0.00\% & 14.40\% & 58.90\% & 42.62\% \\
Kupiec PF (one-sided)      & 8.31\% & 13.93\% & 10.42\% & 20.37\% & 0.00\% & 14.40\% & 8.78\% & 14.64\% \\ \bottomrule
\end{tabular}
\begin{tabular}{lcccc}
\toprule
\multirow{2}{*}{} & \multicolumn{2}{c}{10-day $\text{VaR}_{0.99}$} & \multicolumn{2}{c}{10-day $\text{VaR}_{0.95}$}\\ \cmidrule(lr){2-3} \cmidrule(lr){4-5}
                            & nSDE &  FHS & nSDE  & FHS \\ \cmidrule(lr){1-5}
Coverage ratio median       & 1.0000 & 0.9794 & 0.9671 & 0.9630 \\
Coverage ratio mean         & 0.9949 & 0.9742 & 0.9659 & 0.9367  \\
Kupiec PF (two-sided)      & 0.12\% & 22.72\% & 43.44\% & 37.59\%\\
Kupiec PF (one-sided)      & 0.12\% & 22.72\% & 7.61\% & 20.37\%  \\ \bottomrule
\end{tabular}
\caption{Summary statistics on coverage tests for 2-day, 5-day and 10-day VaRs computed by the neural-SDE (nSDE) market model and the filtered historical simulation (FHS) approach. These results are for DAX options.}
\label{tab:backtesting_mpor_dax}
\end{table}

\begin{figure}[!ht]
    \centering
    \includegraphics[scale=.66]{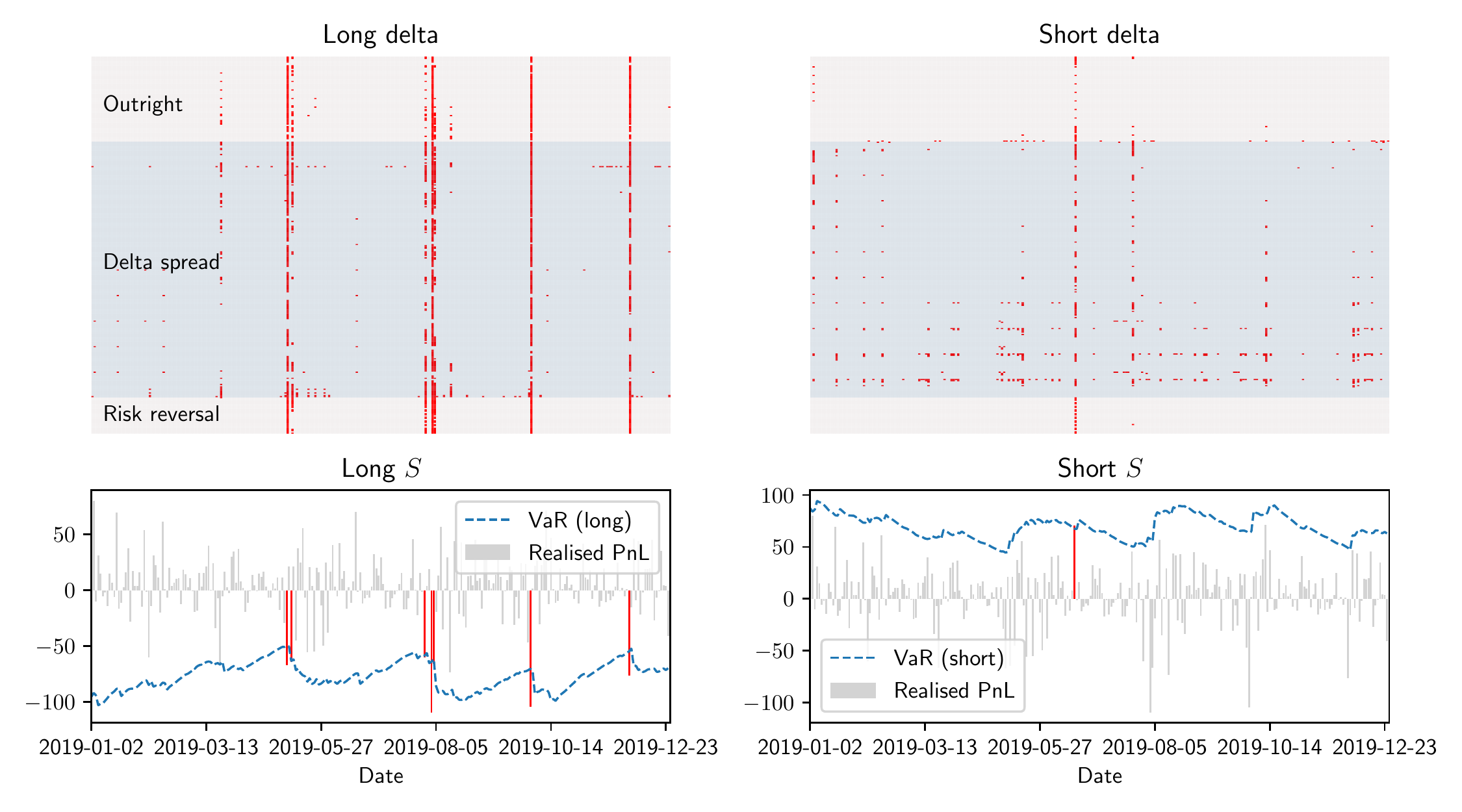}
    \caption{Times series of 1-day FHS-$\text{VaR}_{0.99}$ breaches (red dots) for delta-exposed option portfolios, compared with that for positions in the underlying index.}
    \label{fig:VaR_breach_heatmap_fhs}
\end{figure}

\begin{figure}[!ht]
    \centering
    \includegraphics[scale=.66]{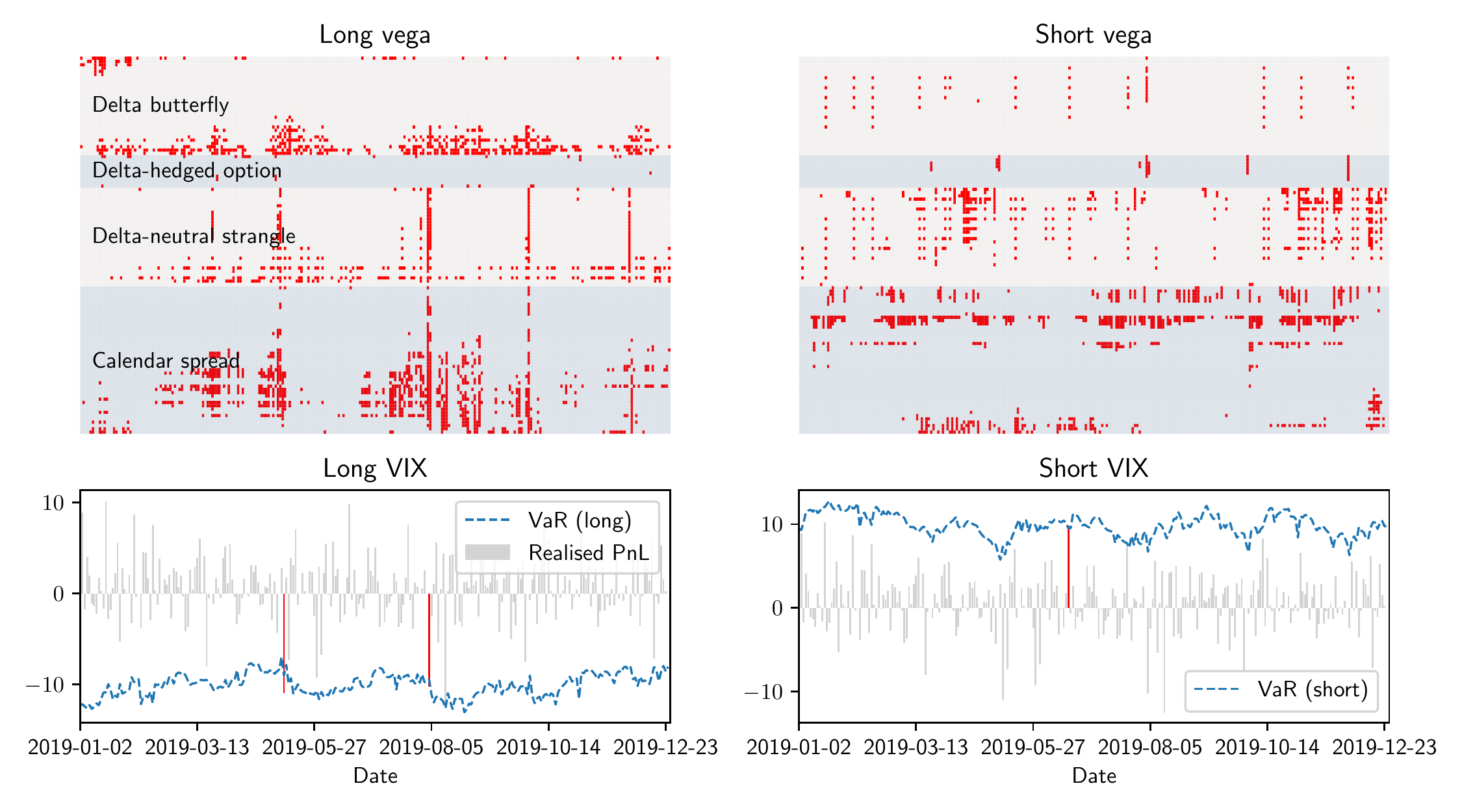}
    \caption{Times series of 1-day FHS-$\text{VaR}_{0.99}$ breaches (red dots) for delta-neutral and vega-exposed option portfolios, compared with that for positions in the VIX portfolio.}
    \label{fig:VaR_breach_heatmap_dn_fhs}
\end{figure}

In Figure \ref{fig:VaR_breach_heatmap_fhs} and \ref{fig:VaR_breach_heatmap_dn_fhs}, we show the time series of 1-day FHS-$\text{VaR}_{0.99}$ breaches for delta-exposed and vega-exposed option portfolios respectively, and compare with those for positions in the underlying index and the VIX portfolio. These are results for EURO STOXX 50 index options, and are supposed to provide a comparison with Figure \ref{fig:VaR_breach_heatmap} and \ref{fig:VaR_breach_heatmap_dn}. The FHS-VaRs tend to have poor coverage performance for delta-hedged option portfolios.

\begin{figure}[!ht]
    \centering
    \includegraphics[scale=.66]{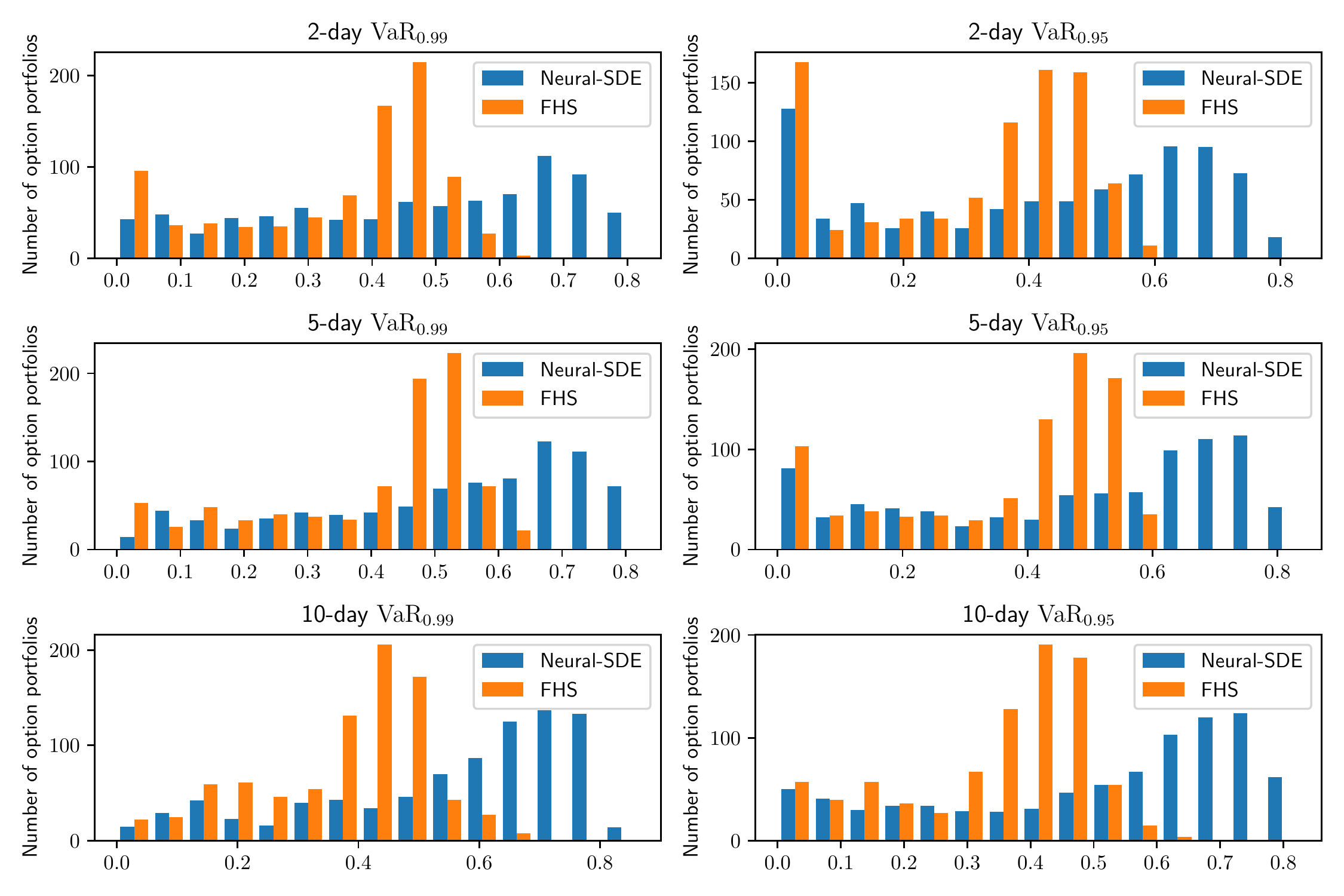}
    \caption{Distribution of trough-to-peak ratios over different tested portfolios of options on the EURO STOXX 50 index.}
    \label{fig:ttp_EX50}
\end{figure}

\begin{figure}[!ht]
    \centering
    \includegraphics[scale=.66]{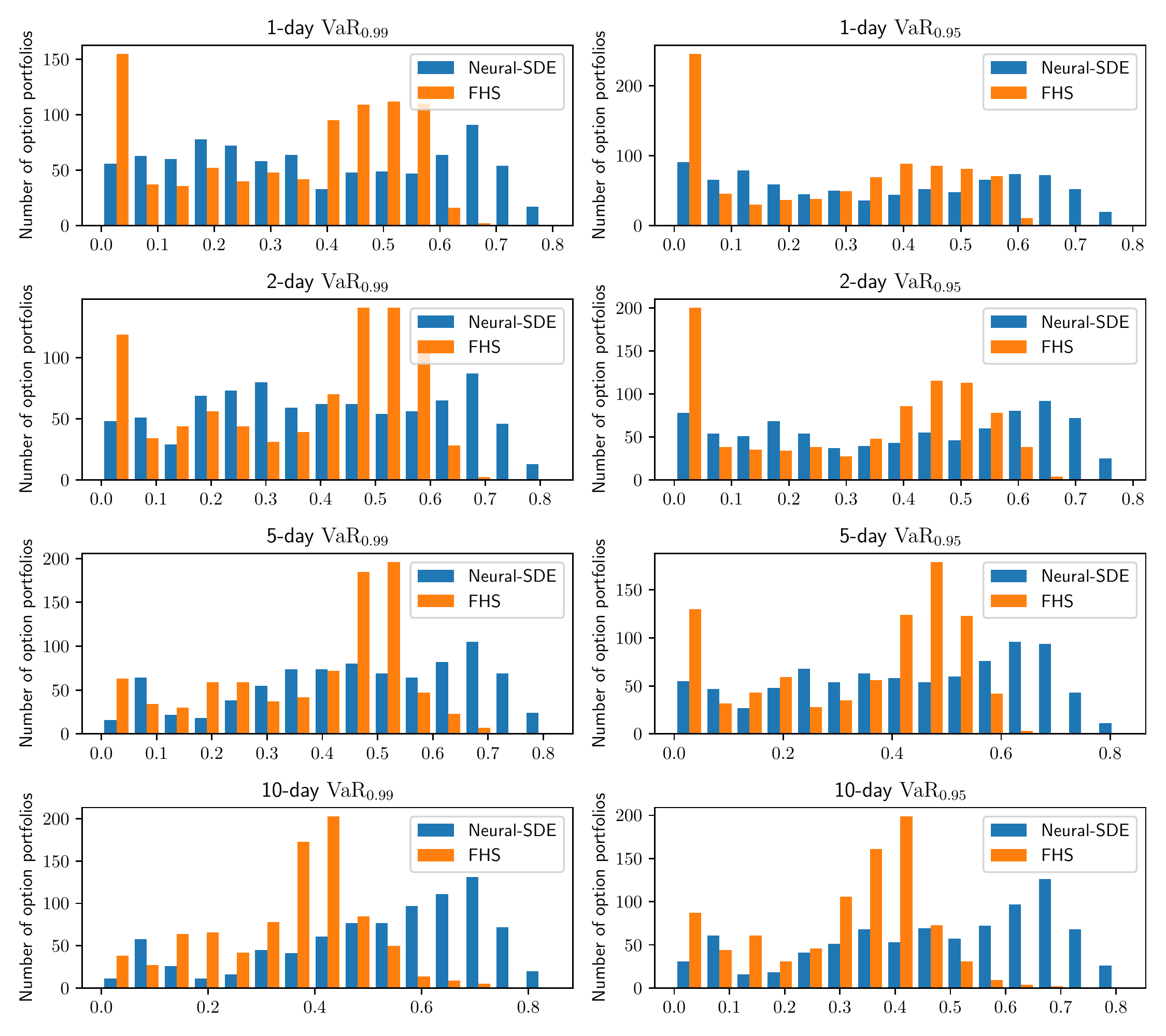}
    \caption{Distribution of trough-to-peak ratios over different tested portfolios of options on the DAX index.}
    \label{fig:ttp_DAX}
\end{figure}

In addition to the coverage and independence test results, in Figure \ref{fig:ttp_EX50} and \ref{fig:ttp_DAX}, we show the comparison of more procyclicality results produced by the nSDE-VaRs and the FHS-VaRs for both index options. The nSDE-VaRs are consistently less procyclical than the FHS-VaRs in all scenarios.

Finally, we compare the VaR backtesting results for EURO STOXX 50 index options generated by the neural-SDE models with different configurations. Specifically, we have described the model with three primary factors (referred to as ``nSDE-3'') in Appendix \ref{apd:3factor} and the model where a full diffusion matrix of $(S, \xi)$ is specified (referred to as ``nSDE-J'') in Appendix \ref{apd:model_xiS}. In Table \ref{tab:backtesting_diffmodels}, we list the statistics on coverage and independence tests for 1-day, 2-day and 5-day VaRs computed by the two models. Compared with the original model for which the related 1-day VaR backtesting results are shown in Table \ref{tab:backtesting_mpor1}, the 3-factor model slightly outperforms the original 2-factor model for most tests, with 17.3\% more portfolios in the green zone. Nevertheless, when compared with the original model for which the related 2-day and 5-day VaR backtesting results are shown in Table \ref{tab:backtesting_mpor2}, neither the 3-factor model nor the original 2-factor model is significantly better than the other.

\begin{table}[!ht]
\scriptsize
\centering
\begin{tabular}{lcccc}
\toprule
\multirow{2}{*}{}                             & \multicolumn{2}{c}{1-day $\text{VaR}_{0.99}$} & \multicolumn{2}{c}{1-day $\text{VaR}_{0.95}$} \\ \cmidrule(lr){2-3} \cmidrule(lr){4-5}
                            & nSDE-3 &  nSDE-J & nSDE-3  & nSDE-J \\ \cmidrule(lr){1-5}
Coverage ratio median       & 0.9881 & 0.9960 & 0.9643 & 0.9742 \\
Coverage ratio mean         & 0.9891 & 0.9911 & 0.9603 & 0.9667 \\
Kupiec PF (two-sided)      & 4.68\% & 11.12\% & 28.81\% & 51.41\% \\
Kupiec PF (one-sided)      & 4.68\% & 11.12\% & 5.15\% & 6.32\% \\ \bottomrule
\end{tabular}
\begin{tabular}{lcccccccc}
\toprule
\multirow{2}{*}{}                             & \multicolumn{2}{c}{2-day $\text{VaR}_{0.99}$} & \multicolumn{2}{c}{2-day $\text{VaR}_{0.95}$} & \multicolumn{2}{c}{5-day $\text{VaR}_{0.99}$} & \multicolumn{2}{c}{5-day $\text{VaR}_{0.95}$} \\ \cmidrule(lr){2-3} \cmidrule(lr){4-5} \cmidrule(lr){6-7} \cmidrule(lr){8-9}
                            & nSDE-3 &  nSDE-J & nSDE-3  & nSDE-J & nSDE-3 & nSDE-J & nSDE-3  & nSDE-J \\ \cmidrule(lr){1-9}
Coverage ratio median       & 0.9960 & 1.0000 & 0.9641 & 0.9880 & 1.0000 & 1.0000 & 0.9637 & 0.9899 \\
Coverage ratio mean         & 0.9946 & 0.9938 & 0.9634 & 0.9718 & 0.9966 & 0.9946 & 0.9694 & 0.9731 \\
Kupiec PF (two-sided)      & 1.05\% & 3.04\% & 33.84\% & 65.11\% & 0.00\% & 3.75\% & 40.52\% & 75.76\% \\
Kupiec PF (one-sided)      & 1.05\% & 3.04\% & 4.57\% & 6.91\% & 0.00\% & 3.75\% & 2.34\% & 8.08\% \\ \bottomrule
\end{tabular}
\caption{Summary statistics on coverage and independence tests for 1-day, 2-day and 5-day VaRs computed by the neural-SDE market models of different configurations. These results are for EURO STOXX 50 options.}
\label{tab:backtesting_diffmodels}
\end{table}

\section*{Acknowledgements}
This publication is based on work supported by the EPSRC Centre for Doctoral Training in Industrially Focused Mathematical Modelling (EP/L015803/1) in collaboration with CME Group.

Samuel Cohen and Christoph Reisinger acknowledge the support of the Oxford-Man Institute for Quantitative Finance, and Samuel Cohen also acknowledges the support of the UKRI Prosperity Partnership Scheme (FAIR) under the EPSRC Grant EP/V056883/1, and the Alan Turing Institute.

\setlength{\bibsep}{0.0pt}
\bibliographystyle{abbrv}
\bibliography{reference}

\end{document}